\documentclass[a4paper,11pt]{article}
\usepackage{pos}
\usepackage{nicefrac}
\usepackage{slashed}

\usepackage{genyoungtabtikz}
\Yboxdim{8pt}
\Ylinethick{1pt}
\newcommand{\md}{\mathrm{d}}

\newcommand{\bone}{{\color{blue}b_{-1}}}
\newcommand{\btwo}{{\color{blue}b_{-2}}}
\newcommand{\boneh}{{\color{blue}b_{-\nicefrac12}}}
\newcommand{\bthreeh}{{\color{blue}b_{-\nicefrac32}}}
\newcommand{\bfiveh}{{\color{blue}b_{-\nicefrac52}}}
\newcommand{\alphaone}{{\color{red}\alpha_{-1}}}
\newcommand{\alphatwo}{{\color{red}\alpha_{-2}}}
\newcommand{\alphathree}{{\color{red}\alpha_{-3}}}
\def\ket#1{\left| #1\right\rangle}

\newcommand{\LeadingB}[1]{%
    {\Ylinecolour{red}\gyoung(#1)}%
    }
\newcommand{\LeadingF}[1]{%
    {\Ylinecolour{red}
    \gyoung(#1)_{\color{red}\nicefrac12}}%
}
\newcommand{\depone}[1]{%
    {\Ylinecolour{yellow}\gyoung(#1)}%
    }

\newcommand{\deptwo}[1]{%
    {\Ylinecolour{blue}\gyoung(#1)}%
    }

\newcommand{\depthr}[1]{%
    {\Ylinecolour{teal}\gyoung(#1)}%
}

\newcommand{\NoGSO}[1]{%
    {\Ylinecolour{gray}\gyoung(#1)}
}

\title{On the deep string spectrum}
\author*[a]{Chrysoula Markou}

\affiliation[a]{Scuola Normale Superiore and INFN,\\
Piazza dei Cavalieri 7, 56126 Pisa, Italy}

\emailAdd{chrysoula.markou@sns.it}

\abstract{These proceedings are based on the author's invited talk reviewing the original published work \cite{Markou:2023ffh, Basile:2024uxn} of the author with collaborators. The subject matter is a new, covariant and efficient technology of constructing entire trajectories of physical string states deeper inside the string spectrum than the leading Regge. The key observation behind the technology is that the lowering operators of a symplectic algebra appear in the Virasoro constraints which impose physicality of states in the open bosonic string. This algebra commutes with the spacetime Lorentz algebra, (of the little group) of which all string states are irreducible representations. Employing then the so--called Howe duality of representation theory, one may relate the irreducible representations of the two algebras via a bijection. The spectrum thus splits into two parts: trajectories that are lowest weight states of the symplectic algebra and their infinitely many clones. The latter can then be reached by suitably dressing the former with the raising operators of the symplectic algebra. The technology is nontrivially extended to the open superstring, where the relevant Howe dual is an orthosymplectic algebra.}

\FullConference{Proceedings of the Corfu Summer Institute 2024 "School and Workshops on Elementary Particle Physics and Gravity" (CORFU2024)\
12 - 26 May, and 25 August - 27 September, 2024\\
Corfu, Greece\\}

\begin{document}
\maketitle

\section{Introduction}

String theories are typically associated with two universal string parameters, the dimensionful string scale $\alpha'$ and the dimensionless string coupling $g_S$, of which the latter is thought of as being determined by the vacuum expectation value of a scalar field, the so--called dilaton, that appears as a massless state in closed string spectra. In string theory there appears thus a single free parameter, namely $\alpha'$. It is precisely $\alpha'$ that sets the mass of all the infinitely many yet \textit{physical} string states of which string spectra consist:
\begin{align}
    M^2=\textrm{integer} \times \frac{1}{\alpha'}\,,
\end{align}
where the factor of proportionality is an integer that enumerates the (infinitely many) mass levels and which depends on the theory in question; for example, for the open bosonic string (in the critical dimension), the integer is $N-1$, with $N=0,1,\dots\,$. All string states have thus on--shell mass and so are thought of as mass eigenstates, while their polarization tensors are transverse and traceless. Consequently, they also correspond to (unitary) irreducible representations of the little group of the spacetime Lorentz group, that is $SO(D-1)$ and $SO(D-2)$ for massive and massless states respectively, $D$ being the number of spacetime dimensions. Intriguingly, because of these properties, one may think of the infinitely many string states as \textit{$1$--particle} states \`a la Bargmann and Wigner. Crucially, among them there appear \textit{infinitely} many massive \textit{higher--spin} states, the appearance of all of which is the reason behind string amplitudes' celebrated UV finiteness.

\section{The rudiments of traditional methods}

Yet what do string spectra look like? Could they be concealing a bigger organizing symmetry? Let us look at the simplest case, the open bosonic string in the critical dimension, namely $D=26$\footnote{The absence of negative--norm states in the bosonic string spectrum has been shown \cite{Kato:1982im, Hwang:1982mc} in this ``critical'' dimension.}. In terms of the Young diagrams that capture the properties of the polarization tensors describing states, the decomposition of the first few levels is given in table \ref{table:one}, see e.g. \cite{Manes:1988gz} and also \cite{Markou:2023ffh}. For now, it suffices to think of a Young diagram as the arrangement of boxes along rows, with each of the boxes representing a spacetime index $\mu=0,1,\dots,D-1$ of a given polarization tensor and each row being totally symmetric under interchanges of its boxes (in the so--called symmetric base), while additional symmetries relate boxes of different rows. For example, a scalar is denoted by a bullet, while a spin--$2$ state, whose polarization  $\varepsilon_{\mu \nu}$ is a symmetric rank--$2$ tensor, can be depicted by the diagram $\gyoung(;;)\,$; we will come back to the systematics of this picture later on. What do we observe in table \ref{table:one}? Starting with the tachyon, the open bosonic string spectrum has a single massless state, which is a vector. The next level contains a massive spin--$2$ state and from level $3$ and on we start seeing massive higher--spins as well as second apparitions of states we've already seen at a lower level: for example, a massive spin--2 state also appears at level $4$. We also observe mixed--symmetry states, such as the hook $\gyoung(;;,;)$ at level $4$. As the level increases,  diagrams with more than $1$ or $2$ rows are possible. If we continue, some diagrams may also appear with a non--trivial multiplicity: for example, there are two massive spin--$2$ states at level $6$, so degenerate states are possible. But how are all these states constructed?

Let us go back to the string field $X^\mu$, which describes the propagation of the string in spacetime. Most of what will be reviewed in the rest of this section can be found, for example, in the classic reference books \cite{Green:1987sp,Polchinski:1998rq,Blumenhagen:2013fgp}, but for a few points  we will refer to the original sources. Solving the equation of motion of $X^\mu$ from the Polyakov action tells us that it can be expanded in modes $\alpha_n^\mu\,$, $n\in \mathbb{Z}\,$, 
\begin{align} \label{expansionb}
     i\partial X^\mu(z) = \sum_{n \in \mathbb Z} \alpha^\mu_n\,z^{-n-1} \quad \Rightarrow \quad     \alpha^\mu_{-n<0} =\oint \frac{\md z}{2 \pi i}\frac{1}{z^n}\,i\partial X^\mu(z)\,,
\end{align}
with  $z$ the complexified worldsheet coordinates and $\partial:=\partial/\partial z \, $ and similarly for its antiholomorphic counterpart. Restricting to the (quantized) open string, the $\alpha_n^\mu\,$, $n\in \mathbb{Z}\,$, satisfy the oscillator algebra
\begin{align} \label{oscalg}
[\alpha^\mu_m,\alpha^\nu_n]= m \delta_{m+n}\eta^{\mu\nu}\,,
\end{align}
where $\alpha_0^\mu:=\sqrt{2\alpha'}p^\mu$, $p^\mu$ being the momentum of the string center of mass, thus implying that a Fock space of string states  can be defined. In particular, the defining property of the vacuum $\ket{0;p}$, namely the momentum eigenstate, is that it be annihilated by all positive modes, namely $\alpha_{k>0}^\mu \ket{0;p}=0$. A number operator can then be defined via
\begin{align}
    N_m^B := \alpha_{-m}\cdot \alpha_m\,,\quad m>0\,,
\end{align}
where $B$ stands for ``bosonic'' and the central dot stands for the spacetime scalar product; $N_m^B$ counts the number of negative modes $m$ excited to construct a given function. Generic string states are functions $F$ of the negative modes $\alpha_{k<0}^\mu\,$ acting on the vacuum $\ket{0;p}$, namely
\begin{align} \label{eq:cand}
    \ket{\textrm{phys}} = g_{\textrm{o}}\, T^a \, F(\alpha_{-1}^\mu, \alpha_{-2}^\nu, \dots) \ket{0;p}\,,
\end{align}
where a priori arbitrary spacetime tensors $\varepsilon_{\mu \nu \dots}$ contract the oscillators' spacetime indices so that string states be spacetime scalars. In the above, $g_{\textrm{o}}$ is the open string coupling, which encodes the strength of the states' interactions and the $T^a$, the so--called Chan--Paton factors, are the generators of the gauge algebra associated with the $D$--branes to which the open string endpoints are thought of as being attached. In the following, we will suppress the factor $g_{\textrm{o}}\, T^a$ from all string states, with its presence always being implicit.

\begin{table}
\centering 
\renewcommand{\arraystretch}{1.5}
  \begin{tabular}{ c || l  }
   $N$ & decomposition in physical states  \\ \hline \hline
   $0$ & $\bullet$ \\
   $1$ & $\gyoung(;)_{so(D-2)}$ \\
   $2$ & $ \deptwo{;;}$ \\
   $3$ & $\gyoung(;;;) \oplus \gyoung(;,;)$ \\
   $4$ & $\gyoung(;;;;) \oplus \gyoung(;;,;) \oplus \deptwo{;;} \oplus \bullet$ \\
   $5$ & $\gyoung(;;;;;) \oplus \gyoung(;;;,;) \oplus \gyoung(;;;) \oplus \gyoung(;;,;) \oplus \gyoung(;,;) \oplus \gyoung(;)$  \\
   $6$ & $\gyoung(;;;;;;) \oplus \gyoung(;;;;,;) \oplus \gyoung(;;;;) \oplus \gyoung(;;;,;) \oplus \gyoung(;;,;) \oplus \gyoung(;;;)  \oplus \gyoung(;;,;;) \oplus \textcolor{red}{2}\, \deptwo{;;} \oplus \gyoung(;,;,;) \oplus \gyoung(;)\oplus \bullet $
  \end{tabular}
\renewcommand{\arraystretch}{1}\caption{Open bosonic critical string, physical content of the first few levels in terms of the respective Young diagrams.} \label{table:one}
\end{table}

As a remnant of the equation of motion of the non--dynamical worldsheet intrinsic metric, physical string states $\ket{\textrm{phys}}$ are those functions of the negative modes that also satisfy the Virasoro constraints \cite{DelGiudice:1970dr}\footnote{The appearance of $\delta_{n,0}$ in the $L_0$ constraint is necessary for the absence of negative--norm states in the critical dimension.}
\begin{align}
    (L_n-\delta_{n,0}) \ket{\textrm{phys}}=0\,, \quad \forall n\geq 0 \,,
\end{align}
where the operators
\begin{align}
      L_n :=\frac{1}{2} \displaystyle \sum_{m=-\infty}^{+\infty} :\alpha_{n-m} \cdot \alpha_m : \,,
\end{align}
generate the famous Virasoro algebra \cite{Fubini:1971ce}
\begin{align}
    [L_m,L_n]=(m-n) L_{m+n}+\frac{c}{12}m(m^2-1)\delta_{m+n,0}\,.
\end{align}
In the above, $c$ is the so--called central charge associated with the conformal field theory (CFT) of the primary field $\partial X^\mu$ on the worldsheet and which is equal to $D$ for the critical string we are considering. In fact, imposing only three constraints, 
\begin{align} \label{eq:vir}
    ( L_0 -1) \ket{\textrm{phys}}=0 \,, \quad  L_1 \ket{\textrm{phys}}=0 \quad , \quad L_2 \ket{\textrm{phys}}=0\,,
\end{align}
is sufficient, see for example \cite{Sasaki:1985py}, since all others follow by applying the Virasoro algebra. Moreover, defining the level $N$ and mass $M$ of a given state via
\begin{align}
    N:= \sum_{m=1} N_m^B =0,1,2,\dots \,,\quad M^2:=-p^2\,,
\end{align}
the $L_0$ constraint of \eqref{eq:vir} takes the form
\begin{align}
    M^2= \frac{N-1}{\alpha'}\,,
\end{align}
namely yields the infinite mass spectrum of the open bosonic string.

So how can this knowledge be employed to construct table \ref{table:one}? There are several ways, each with its own advantage and limitation; essentially all proceed on a level--by--level basis. Let's first consider the ``old covariant way''. Its recipe stands as follows: first, choose a value of the level $N$ and write the respective Ansatz for the polynomial $F$ of the string states \eqref{eq:cand} at the level in question. Next, substitute the Ansatz into and solve the constraints \eqref{eq:vir} imposed by $L_1$ and $L_2$. The output will be the physical states at the chosen level. Let us work out the first few levels as examples. At $N=0$, there are no oscillators excited so $F=1$ and the level contains a single tachyonic state 
\begin{align}
     \ket{\textrm{tachyon}} = \ket{0;p}\,,\quad M^2=-1/\alpha'\,,
\end{align}
namely the first scalar we see in table \ref{table:one}. At $N=1$, only the oscillator $\alpha_{-1}^\mu$ can be excited, so the Ansatz takes the form $F=\varepsilon\cdot \alpha_{-1}$, where $\varepsilon^\mu$ is an a priori arbitrary spacetime vector. $L_1$ imposes that $\varepsilon^\mu$ be transverse and $L_2$ imposes no additional constraints: this level contains a single massless vector
\begin{align}
    \ket{\textrm{vector}}=  \varepsilon \cdot \alpha_{-1} \, \ket{0;p}  \,,\quad p \cdot  \varepsilon=0\,,\quad M^2=0\,,
\end{align}
the \textit{only} massless state of the open bosonic string. At level two, there are two possible terms in the Ansatz: $F= \varepsilon_{\mu \nu} \, \alpha_{-1}^{\mu} \, \alpha_{-1}^{\nu} + \tilde{\varepsilon} \cdot \alpha_{-2} $, where $ \varepsilon_{\mu \nu}$ and $\tilde{\varepsilon}^\mu$ are a symmetric rank--2 tensor and a vector respectively, on which $L_1$ and $L_2$ impose then two algebraic constraints. However, the trace part of $ \varepsilon_{\mu \nu}$ as well as the vector $ \tilde{\varepsilon}_\mu$ are actually null states and $L_1$ and $L_2$ imply then that level contains a single massive spin--$2$ state with a transverse and traceless polarization tensor \cite{Friedan:1985ge}
\begin{align}
\ket{\textrm{phys}}=\varepsilon_{\mu \nu} \, \alpha_{-1}^{\mu} \, \alpha_{-1}^{\nu}  \ket{0;p}\,,\quad p^\nu \varepsilon_{\mu \nu} = 0 \,,\quad \varepsilon_\mu{}^\mu = 0 \,\quad M^2=1/\alpha'\,.
\end{align}

As the level increases, its possible partitions also increase in number, so that the size of the Ansatz increases and solving the $L_1$ and $L_2$ constraints becomes a harder and harder problem. There is, however, a set of infinite states for which it is a simple exercise to write the respective physical polynomial: these are the highest--spin states per level, whose polarization tensors are symmetric rank--$s$ tensors and which comprise the so--called ``leading Regge trajectory'' of states. More specifically, let us consider an arbitrary level $N=s$, $s \in \mathbb{N}$. Since constructing physical states at this level essentially amounts to suitably distributing $s$ units of energy among the oscillators $\alpha_{-1}^\mu\,,\alpha_{-2}^\mu,\dots$ in all possible ways, the way to distribute the units to as many oscillators as possible without changing the level is to use $s$ oscillators of the type $\alpha_{-1}^\mu$:
\begin{align} \label{leading}
\ket{\textrm{leading}}:=F_1   \ket{0;p}:=\varepsilon_{\mu_1 \dots \mu_s} \, \alpha_{-1}^{\mu_1} \dots \alpha_{-1}^{\mu_s}  \ket{0;p}\,,
\end{align}
where $\varepsilon_{\mu_1 \dots \mu_s}$ is an a priori arbitrary symmetric rank--$s$ tensor. Indeed, this trajectory is physical if $\varepsilon_{\mu_1 \dots \mu_s}$ is transverse and traceless, since
\begin{align} \label{eq:virleadingt1}
L_1 F_1  \ket{0;p} &\overset{!}{=}0 \quad \Rightarrow \quad ( p\cdot \alpha_{1}+\ldots) F_1 \ket{0;p} =0\quad \Rightarrow \quad  p^\nu \varepsilon_{\nu \mu_2...\mu_s}=0\,, \\ \label{eq:virleadingt2}
L_2 F_1  \ket{0;p} &\overset{!}{=}0 \quad \Rightarrow \quad(\alpha_1 \cdot \alpha_1 +\ldots) F_1 \ket{0;p} =0\quad  \Rightarrow \quad \eta^{\nu\sigma}\varepsilon_{\nu \sigma \mu_3...\mu_s}=0\,,
\end{align}
where the dots stand for terms containing annihilation oscillators that carry more than $1$ unit of energy, which commute with $\alpha_{-1}^\mu$ and so do not contribute to the constraints. The first few states of this trajectory are highlighted in red in table \ref{table:two} and the tachyon, massless vector and lightest massive spin-2, whose construction we previously reviewed on a level--by--level basis, are its lightest member--states. It is worth noting that the expansion \eqref{expansionb} yields the dictionary
\begin{align} \label{dictionaryb}
   \ket{p;0} = \lim_{z\to0} e^{i p \cdot X(z)} \ket0 \,,\quad   \alpha^\mu_{-n<0} \ket0    \,\leftrightarrow \,  \frac{1}{(n-1)!}\, i\partial^n X^\mu(0)  \,,
\end{align}
which can be used to map string states, as functions of the oscillators they excite, to the respective vertex operators in the CFT of the primary field $\partial X$, essentially by replacing the vacuum state by a plane wave and oscillators with descendants of conformal weight equal to the units of energy the former carry, in the context of the state--operator correspondence. The Virasoro constraints are then formulated as imposing that the worldsheet BRST charge commutes with physical vertex operators up to total derivatives and, for the case of the leading Regge trajectory, the constraints \eqref{eq:virleadingt1}--\eqref{eq:virleadingt2} can be written as differential constraints on the function $F_1$ \cite{Sagnotti:2010at}:
\begin{align}
    p \cdot \frac{\delta F_1}{\delta \partial X}=0   \,, \quad   \frac{\delta^2 F_1}{\delta \partial X  \cdot \delta \partial X} =0\,.
\end{align}

In part because of the complication of solving the Virasoro constraints as the level increases, there exist alternatives, each with its own limitation. In the so--called light--cone construction of the spectrum \cite{Goddard:1973qh}, a ``light--cone'' gauge is chosen, in which the Virasoro constraints become linear and string states can be written as functions of the transverse oscillators $ \alpha^{i}_{-n} \,$, $i=0,1,\dots,D-2$. As such, they are immediately physical, but fall into irreducible representations of $SO(D-2)$; to recombine them into $SO(D-1)$ irreducible representations so as to recognize them as irreducible representations of the spacetime little group is again a problem that becomes harder and harder as the level increases. An alternative is the powerful construction of the spectrum \'a la Del Giudice, Di Vecchia, Fubini \cite{DelGiudice:1971yjh}, in which the vertex operator of an excited string state is the scattering amplitude of a number of photons, corresponding appropriately to the types of oscillators excited in the state, off an initial tachyon. However, the latter's momentum enters the framework as a reference vector which appear in all physical polarizations, eliminating which inevitably brings back the Virasoro constraints and their associated difficulty to solve at a high level. Finally, a partition function is available \cite{Curtright:1986di,Curtright:1986rr,Hanany:2010da}, from which the characters of the physical states can be read in principle at any level. However, constructing the $SO(D-1)$ irreducible representations from them as well as the respective states in terms of the oscillators they excite is again a problem that becomes harder with the level. It should be highlighted that all the aforementioned techniques are equivalent and produce physical string states which, once written in terms of the little group's irreps, have polarizations that are transverse and traceless (TT).

\begin{table}
\centering 
\renewcommand{\arraystretch}{1.5}
  \begin{tabular}{ c || l  }
   $N$ & decomposition in physical states  \\ \hline \hline
   $0$ & $\textcolor{red}{\bullet}$ \\
   $1$ & $\LeadingB{;}_{\textcolor{red}{so(D-2)}}$ \\
   $2$ & $ \LeadingB{;;}$ \\
   $3$ & $\LeadingB{;;;} \oplus \gyoung(;,;)$ \\
   $4$ & $\LeadingB{;;;;} \oplus \gyoung(;;,;) \oplus \deptwo{;;} \oplus \bullet$ \\
   $5$ & $\LeadingB{;;;;;} \oplus \gyoung(;;;,;) \oplus \deptwo{;;;} \oplus \gyoung(;;,;) \oplus \gyoung(;,;) \oplus \gyoung(;)$  \\
   $6$ & $\LeadingB{;;;;;;} \oplus \gyoung(;;;;,;) \oplus \deptwo{;;;;} \oplus \gyoung(;;;,;) \oplus \gyoung(;;,;) \oplus \gyoung(;;;)  \oplus \gyoung(;;,;;) \oplus 2\, \gyoung(;;) \oplus \gyoung(;,;,;) \oplus \gyoung(;)\oplus \bullet $
  \end{tabular}
\renewcommand{\arraystretch}{1}\caption{Open bosonic critical string, leading Regge trajectory and its first clone highlighted in red and blue respectively.} \label{table:two}
\end{table}

\section{A new technology}

In light of these considerations, the main challenge surrounding the construction of the string spectrum is: what do excited string states look like? This is precisely the question that the work reviewed in the present proceedings aims to tackle and, in this section in particular, we will be reviewing \cite{Markou:2023ffh}. Firstly, there is an observation that can already be made by inspecting table \ref{table:two} more closely, without employing other tools. In particular, we see the beginning of a trajectory that looks like a clone of the leading Regge trajectory truncated right after the massless vector. After this truncation, it contains the same types of Young diagrams but at higher levels: the spin--$2$ appears at level $4$, the spin--$3$ at level $5$ and so on, instead of these states appearing at level $2$ and $3$ respectively. The leading Regge trajectory has namely a truncated clone, highlighted in blue in table \ref{table:two}, each member--state of which appears \textit{two} levels higher than the member--state of the leading Regge with the same Young diagram. Likewise, the $2$--row trajectory that starts at level $3$ with the antisymmetric rank--2 tensor has two clones that start at level $5$: one with the hook and another with the antisymmetric rank--2 tensor as lightest member--states; it is easy to notice further examples. The spectrum thus seems \textit{repetitive}, so a natural question raises itself: is there a certain pattern? Can it be that the spectrum is concealing a bigger organizing symmetry?

To answer this question, let us first get back to the subject of what a general string state looks like and, in particular, what its Young diagram looks like. The spacetime symmetries of \textit{any} string state are encoded in its polarization tensor $\varepsilon^{\mu(s_1),\, \lambda (s_2), \dots,\, \nu(s_K)}\,(p) $, that is a tensor of $\mathfrak{gl}$, which can be depicted by a Young diagram as
\begin{align} \label{eq:yds}
 \Yboxdim{14pt} \gyoung(_9{s_1},_8{s_2},_6{\dots},_4{s_K}) \quad  \Leftrightarrow \quad \varepsilon^{\mu(s_1),\, \lambda (s_2), \dots,\, \nu(s_K)}\,(p) \,.
\end{align}
A few comments on notation and conventions are in order. We are using the symmetric base for Young diagrams, namely a group of $s_1$ symmetric indices $\mu$, denoted as $\mu(s_1):=\mu_1\dots \mu_{s_1}$ in the polarization tensor, is depicted by a row of $s_1$ boxes in the diagram; different groups of symmetric indices are separated by commas and correspond to different rows. The lengths $s_i$ of the $K$ rows, namely the numbers of boxes the latter contain, which are indicated inside the rows of the Young diagram in \eqref{eq:yds}, can never be increasing as we follow the diagram along downwards, namely $s_1 \geq  s_2 \geq \ldots \geq s_K$. Irreducibility under $\mathfrak{gl}$ takes then the form of the Young symmetry condition, which, for the simple example of the ``hook'' $ \gyoung(;;,;) \,\, \Leftrightarrow \,\,  \varepsilon^{\mu_1 \mu_2,\lambda} \,$ reads
\begin{align}
\varepsilon^{\mu_1 \mu_2,\lambda} + \varepsilon^{ \mu_2  \lambda , \mu_1} + \varepsilon^{\lambda \mu_1, \mu_2} =0\,,
\end{align}
namely imposes that fully symmetric permutations of the three indices of the hook's tensor must vanish. Imposing then that irreducible $\mathfrak{gl}(D)$ tensors be transverse and traceless turns them into irreducible $\mathfrak{so}(D-1)$ tensors. For more details on polarization tensors and Young diagrams (in field theory), the reader is referred to \cite{Bekaert:2006py,Didenko:2014dwa}. Crucially, (in string theory) there are \textit{infinitely} many ways of contracting a given (TT) polarization with a polynomial in the oscillators $\alpha_{-1}^\mu,\alpha_{-2}^\nu\, \dots\,$, all of which yield physical string states, albeit at different levels (up to multiplicity). In other words, there are infinitely many ways of embedding a given Young diagram in the string spectrum.

So, what is the \textit{simplest} polynomial with which a general Young diagram can be dressed? The recipe is to contract all boxes of the $i$--th row with $\alpha_{-i}$ \cite{Weinberg:1985tv}. In our notation, this amounts to constructing the polynomials
\begin{align} \label{simplest}
    F_{\textrm{simpl}} = \varepsilon_{\mu(s_1),\, \lambda (s_2), \dots,\, \nu(s_K)} \, \alpha_{-1}^{\mu_1} \dots \alpha_{-1}^{\mu_{s_1}} \,\alpha_{-2}^{\lambda_1} \dots \alpha_{-2}^{\lambda_{s_2}}  \dots \alpha_{-K}^{\nu_1} \dots \alpha_{-K}^{\nu_{s_K}} \,.
\end{align}
By construction, the polarization tensor $\varepsilon_{\mu(s_1),\, \lambda (s_2), \dots,\, \nu(s_K)}$  enjoys Young symmetry and we may further take it to be TT. The polynomial \eqref{simplest} passes then the Virasoro constraints $L_1$ and $L_2$ and, recalling that each $\alpha_{-i}$ carries $i$ units of energy, it is easy to calculate the respective level,
\begin{align}
 N_{\text{min}}=\sum_{i=1}^K s_i \,i \,,
\end{align}
which is clearly the \textit{lowest possible} level in the spectrum in which the general diagram \eqref{eq:yds} can be embedded:
 For example, the simplest polynomial for the hook is 
\begin{align} \label{hookzero}
    F_{\textrm{simpl}}^{  \Yboxdim{4pt} \gyoung(;;,;)} = \varepsilon_{\mu_1 \mu_2,\, \lambda } \, \alpha_{-1}^{\mu_1}  \alpha_{-1}^{\mu_{2}}  \alpha_{-2}^{\lambda}\,,
\end{align}
which appears at level $ N_{\text{min}}=4$, which is precisely what we also see in table \ref{table:one}. Another example is the entire leading Regge trajectory \eqref{leading}, namely the set of $1$--row diagrams of spin--$s$ at level $ N_{\text{min}}=s\,$. It is clear that \textit{any} diagram  appears at its respective $N_{\text{min}}$ for the first time. 

What about the subsequent apparitions of a given Young diagram? Let us begin ``parametrizing our ignorance''. First, a subsequent apparition evidently finds itself at a certain level $N$, that is a certain number of units of energy, say $w$, higher than $N_{\text{min}}$, namely
\begin{align}
    w:= N- N_{\text{min}}\,.
\end{align}
The integer $w$ thus parametrizes how ``deep'' inside the spectrum a given physical state appears with respect to the first apparition of its corresponding Young diagram, so a suggestive name for this parameter can be \textit{depth}. By this definition, the entire class of simplest polynomials \eqref{simplest}, including for example the leading Regge trajectory, appears at $w=0\,$. Next, since at $w=0$ every row of a given Young diagram is associated with the excitation of a \textit{different} oscillator, it appears meaningful to define a \textit{trajectory} as the set of physical states whose Young diagrams have a fixed number of rows and which appear at a fixed value of the depth $w$, bearing in mind that a clear definition has appeared only for the leading Regge trajectory in past literature. For example, the previously referred to clone of the leading Regge trajectory, which is highlighted in blue in table \ref{table:two}, is a $1$--row trajectory at depth $w=2\,$. But what does its polynomial look like?

Before considering the answer to this question, let us review how the spectrum can now be re--organized in terms of the depth $w$, instead of the level $N$. This is reflected in table \ref{table:three}, where all (lightest member--states of) trajectories at depths $w=0,1,2,3$ are highlighted in red, yellow, blue and teal respectively. Several comments are in order. First, it is clear that the string spectrum consists of \textit{two parts}: all possible trajectories at depth $w=0$, whose polynomials are \textit{known}, and all their \textit{clones} (w.r.t. the spacetime symmetries of the respective polarization tensors) at $w>0$. The polynomials of the latter is what we will be looking for in a bit, but a measure of their complexity is undoubtedly the depth $w$. In addition, a given trajectory may have several branches, that start at different levels. For example, the $2$--row trajectory at $w=0$ has a branch that starts at level $3$ with the antisymmetric tensor $\gyoung(;,;)\,$, another that starts at level $6$ with the ``window'' $\gyoung(;;,;;)\,$, as well as several others. This is simply because not all diagrams can appear at the same level for their $n$--th apparition, since every time a new box is added to a fixed diagram at a fixed level, the level increases. Also, we clearly do not see the lightest member--states of all possible trajectories in table \ref{table:three}: for example, the lightest member--state of the $4$--row trajectory at $w=0$ is a rank--$4$ antisymmetric tensor $\gyoung(;,;,;,;)$, for which the lowest possible level is $N_{\text{min}}=10\,$. In other words, as the level increases, more trajectories start appearing as well as more clones. Each trajectory has infinitely many clones, some of which may be truncated. And lastly, it is worth stressing that along a given trajectory, arbitrarily high spin can be reached for a finite number of types of oscillators excited, simply by adding arbitrarily many boxes to the rows of the trajectory. For example, along the leading Regge, arbitrarily high spin is reached by exciting arbitrarily many copies of the oscillator $\alpha_{-1}$, without exciting other types of oscillators. This will also be the case for trajectories at $w>0$, as we will see.
\begin{table}
\centering 
\renewcommand{\arraystretch}{1.5}
  \begin{tabular}{ c || l  }
   $N$ & decomposition in physical states \hspace{5.5cm}  $w=\textcolor{red}{0}, \textcolor{yellow}{1}, \textcolor{blue}{2}, \textcolor{teal}{3}, \dots$ \\ \hline \hline
   $0$ & $\textcolor{red}{\bullet}$ \\
   $1$ & $\textcolor{red}{\LeadingB{;}_{so(D-2)}}$ \\
   $2$ & $ \LeadingB{;;}$ \\
   $3$ & $\LeadingB{;;;} \oplus \LeadingB{;,;}$ \\
   $4$ & $\LeadingB{;;;;} \oplus \LeadingB{;;,;} \oplus \deptwo{;;} \oplus \bullet$ \\
   $5$ & $\LeadingB{;;;;;} \oplus \LeadingB{;;;,;} \oplus \depone{;;,;}  \oplus \deptwo{;;;}\oplus \deptwo{;,;} \oplus \gyoung(;)$  \\
   $6$ & $\LeadingB{;;;;;;} \oplus \LeadingB{;;;;,;}  \oplus \LeadingB{;;,;;} \oplus \LeadingB{;,;,;} \oplus \depone{;;;,;} \oplus \deptwo{;;;;} \oplus \deptwo{;;,;} \oplus \depthr{;;;}   \oplus 2\, \gyoung(;;)    \oplus \gyoung(;)\oplus \bullet $
  \end{tabular}
\renewcommand{\arraystretch}{1}\caption{Open bosonic critical string spectrum, with different color shades corresponding to trajectories at different depths up to $w=3$.} \label{table:three}
\end{table}

To systematize the parametrization of the spectrum via the integer $w$, let us consider the most \textit{general} string state, $F(\alpha_{-1}^\kappa, \alpha_{-2}^\nu, \dots) \ket{0;p}$, where an a priori arbitrary polarization tensor $\varepsilon_{\mu(s_1),\lambda(s_2),\dots}$ enters the polynomial $F$, \textit{without} any restriction to a given state or trajectory (or on level/depth) and let us examine the form of the Virasoro constraints acting on it. We will use the transverse subspace simplification \cite{Scherk:1974jj,Manes:1988gz}, namely restrict to
\begin{align} \label{trsub}
    \eta^{\mu \nu}_\perp=\eta^{\mu \nu} - \frac{p^\mu p^\nu}{p^2}\,,
\end{align}
since it is sufficient to consider only \textit{transverse} polarizations to construct all physical string states at a given level, hence the entire spectrum, but the full proof is equally possible without this restriction. What the transverse subspace simplification \eqref{trsub} does is remove the explicit $p$--dependence of the polynomials and replace $D$ with $D-1$, thereby essentially removing all degrees of freedom that do not propagate, namely all longitudinal modes. For pedagogical reasons, we focus on the description in the oscillator language, but the original technology was also developed in the completely equivalent CFT language of the primary field $\partial X$, in which the simplification removes all BRST--exact states, since the (worldsheet) BRST cohomology is in the transverse subspace \cite{Kato:1982im,Henneaux:1986kp}. In particular, having the Virasoro generators act on a general $F$ brings the constraints to the form
\begin{align} \label{virg1}
   (L_0-1)F &= \bigg[ \displaystyle \sum_{m=1}^{\infty} \alpha_{-m}\cdot \alpha_m + \alpha'p^2 -1\bigg]F =0 \\ \label{virg2}
   2L_{n>0}^\perp F& =\bigg[  \displaystyle \sum_{m=1}^{n-1} \alpha_{n-m}\cdot \alpha_m +2  \sum_{m=1}^\infty \alpha_{-m} \cdot \alpha_{n+m} \bigg]F =0\,,
\end{align}
where the scalar products are taken w.r.t. the transverse metric \eqref{trsub} and we recall that the constraints due to $L_0$, $L_1$ and $L_2$ are sufficient. What are the spacetime scalars, that are also oscillator bilinears, of which it is easy now to see that the Virasoro constraints are linear combinations?

Let us construct all possible oscillator bilinears that are spacetime scalars:
\begin{align} \label{opT}
    T^k{}_{\ell}  := \tfrac1k \, \alpha_{-k} \cdot  \alpha_{\ell} \, , \quad  T_{k\ell} :=  \alpha_k \cdot \alpha_{\ell} \, , \quad T^{k\ell}:=\tfrac{1}{k\ell} \, \alpha_{-k}\cdot \alpha_{-\ell}\,,
\end{align}
where $k,\ell=1,2,\dots\,N$, with $N$ being an integer, and the prefactors have been chosen for convenience. The operators $T^k{}_{\ell} \,$, $T_{k\ell} $ and $T^{k\ell} $ carry $k-\ell\,$, $-k-\ell$ and $k+\ell$ units of energy respectively. Employing the oscillator algebra \eqref{oscalg}, it can be shown that the $T$ operators \eqref{opT} generate the algebra
\begin{align} \label{sp1}
       [T^\ell{}_n,T^{km}]&= \delta^k_n T^{\ell m}+\delta^m_n T^{\ell k}\\ \label{sp2}
     [T_{km}, T^\ell{}_n]&= \delta^\ell_k T_{mn}+\delta^\ell_m T_{kn}\\ \label{sp3}
     [T^k{}_\ell,T^m{}_n]&= \delta_\ell^m T^k{}_n-\delta^k_n T^m{}_\ell\\ \label{sp4}
    [T_{km}, T^{\ell n}]&=(D-1)(\delta_k^n \delta_m^\ell + \delta_k^\ell \delta_m^n)+  \delta^\ell_k T^n{}_m+\delta^\ell_m T^n{}_k+\delta^n_k T^\ell{}_m+\delta^n_m T^\ell{}_k\,,
\end{align}
namely (after the redefinition $T^k{}_\ell \rightarrow T^k{}_\ell + \tfrac{D-1}{2}  \delta^k_\ell\,$, to absorb the central term in \eqref{sp4})  the symplectic algebra $\mathfrak{sp}(2N)$! If one considers the complete string spectrum, namely the complete set of oscillators, $N$ is unbounded. For finite $N$, one may split the $\mathfrak{sp}$ generators into raising and lowering operators, and, for example, the following choice may be made
\begin{align}
    \textrm{lowering}&:\quad T_{k\ell}\,,\quad T^{k<\ell}{}_{\ell} \\
 \textrm{raising}&:\quad T^{k\ell}\,,\quad T^{k>\ell}{}_{\ell}\,.
\end{align}
This is a convenient choice, as now the raising and lowering operators $T$ add and remove units of energy respectively. In accordance with this choice, the lowest weight states $\widetilde{F}$ of the $\mathfrak{sp}$ algebra satisfy the conditions
\begin{align}
 T_{k\ell} \widetilde{F}=0\,,\quad T^{k<\ell}{}_{\ell} \widetilde{F}=0  \,,\quad    T^{k}{}_{k} \widetilde{F}=s_k \,\widetilde{F}\,,
\end{align}
namely are annihilated by all $\mathfrak{sp}$ lowering operators, while the eigenvalues of the Cartan subalgebra generators $T^{k}{}_{k}$ are the lengths $s_k$ of the rows of the respective Young diagram.

Which string states are the lowest weight states $\widetilde{F}$ of the $\mathfrak{sp}$? Let us consider the $w=0$ trajectories of the spectrum and, for example, the hook \eqref{hookzero}. It is easy to see that, among the $\mathfrak{sp}$ lowering operators, only the action of $T^1{}_2$, $T_{11}$ and $T_{12}$ on $   F_{\textrm{simpl}}^{  \Yboxdim{4pt} \gyoung(;;,;)} $ yields a non--trivial result and in particular
\begin{align}
T^1{}_2  F_{\textrm{simpl}}^{  \Yboxdim{4pt} \gyoung(;;,;)}&= 0 \quad  \Leftrightarrow \quad \varepsilon^{\mu_1 \mu_2,\lambda} + \varepsilon^{ \mu_2  \lambda , \mu_1} + \varepsilon^{\lambda \mu_1, \mu_2} =0 \\
T_{11} F_{\textrm{simpl}}^{  \Yboxdim{4pt} \gyoung(;;,;)}&=0 \quad \Leftrightarrow \quad \eta^{\mu_1 \mu_2} \varepsilon_{\mu_1 \mu_2, \lambda}=0 \\
 T_{12} F_{\textrm{simpl}}^{  \Yboxdim{4pt} \gyoung(;;,;)}&=0 \quad \Leftrightarrow \quad \eta^{\mu_2\lambda} \varepsilon_{\mu_1 \mu_2, \lambda}=0\,,
\end{align}
while
\begin{align}
   T^1{}_1  F_{\textrm{simpl}} =2   F_{\textrm{simpl}}\,,\quad    T^2{}_2  F_{\textrm{simpl}} =   F_{\textrm{simpl}}\,.
\end{align}
This means that $T^1{}_2$ and $T_{11}$, $T_{12}$ impose respectively the Young symmetry and tracelessness of $\varepsilon^{\mu_1 \mu_2,\lambda}$ upon acting on $F_{\textrm{simpl}}^{  \Yboxdim{4pt} \gyoung(;;,;)}$. More generally, the $\mathfrak{sp}$ lowering operators check the Young symmetry and tracelessness of the polarizations that contract the oscillators of $w=0$ trajectories. In other words, the lowest weights states of the $\mathfrak{sp}$ algebra \textit{are} the $w=0$ trajectories! Turning back to the Virasoro constraints \eqref{virg1}--\eqref{virg2} on a general $F$, these can now be rewritten as
\begin{align} \label{virfin}
   (L_0-1)F &= \bigg[ \displaystyle \sum_{m=1}^{\infty} nT^n{}_n + \alpha'p^2 -1\bigg]F =0 \\ \label{virfin2}
   2L_{n>0}^\perp F& =\bigg[ \displaystyle \sum_{m=1}^{n-1}  \,T_{m,n-m} +2 \sum_{m=1}^\infty mT^m{}_{n+m} \bigg]F =0\,,
\end{align}
using the definitions \eqref{opT} of the $\mathfrak{sp}$ generators. It is now clear that $L_0$ and $L_1$, $L_2$ contain the Cartan (subalgebra of the $\mathfrak{sp}$) generators and the $\mathfrak{sp}$ \textit{lowering} operators respectively. When $F$ corresponds to an $w=0$ state or trajectory, it passes the Virasoro constraints since it is annihilated by all $\mathfrak{sp}$ lowering operators, while $L_0$ embeds it at the correct level. But what about the $w>0$ trajectories?

The answer is tied to the use of an interesting result from representation theory, namely ``\textit{Howe duality}'' \cite{Howe1,Howe2} (see also \cite{Rowe:2012ym,Basile:2020gqi} for pedagogical presentations and various examples). Loosely speaking, is can be phrased as follows: when two algebras commute, in which case they are called ``Howe duals'', then every irreducible representation of one of the two is mapped to a unique irreducible representation of the other and vice versa. The dimensionality of the irreps of the former is then identified with the multiplicity of the irreps of the latter and vice versa. More formally, two algebras are Howe duals when they are subalgebras of the same (symplectic, orthogonal or orthosymplectic) algebra  and each other's centralizers; in that case, the irreps of the two algebras are mapped to each other via a bijection\footnote{More precisely, the bijection relates those irreps that appear in the Fock space, namely in the oscillator realization, of the larger algebra of which the dual pair are subalgebras.}. This is precisely the case for the $\mathfrak{sp}(2N)$ algebra we have just seen and the spacetime Lorentz algebra $\mathfrak{so}(D-1,1)$, of the little group of which (the polarizations of) all physical string states are irreducible representations. In particular, the symplectic and Lorentz algebras act on different types of indices: the former on the modes $n,k,\dots$ and the latter on the spacetime indices $\mu, \nu,\dots$ of the string oscillators. Consequently, Howe duality may be employed in the following way: to every Lorentz irrep there corresponds a unique $\mathfrak{sp}$ irrep (and vice versa). The first apparition of every diagram in the string spectrum corresponds to the lowest weight state of the respective $\mathfrak{sp}$ irrep, namely a $w=0$ string state. The subsequent apparitions can be reached by moving along the respective $\mathfrak{sp}$ module, namely by acting with the $\mathfrak{sp}$ \textit{raising} operators on the lowest weight state.  In other words, starting from a physical trajectory with polynomial $F_{w=0}$ defined by a given number of rows at depth $w=0$, its clone at $w>0$ can be reached by dressing $F_{w=0}$ with a function $f$  of the $\mathfrak{sp}$
raising operators, such that $f$ carries a total  of units of energy equal to the depth $w$ we aim to probe. We thus have the following Ansatz
\begin{align} \label{ansatz}
    F_{w>0} =  f_{w>0}( T^{mn}, T^{k>\ell}{}_{\ell}) \, F_{w=0}\,,
\end{align}
where $f$ contains all possible combinations of the raising operators $T^{mn}, T^{k>\ell}{}_{\ell}$ that carry $w$ units of energy. By construction, $f$ does not modify the spacetime symmetries of the trajectory at $w=0$, as the symplectic generators leave the respective Young diagrams invariant; rather, $f$ increases the level or depth of each of the respective string states, hence the cloning of the entire trajectory is possible. However, of the Ans\"atze \eqref{ansatz}, only the subset that passes the Virasoro constraints is physical, the $\mathfrak{sp}$ thus \textit{not} being a symmetry of the string spectrum. It is also worth mentioning that, for a fixed depth, the Ansatz clearly contains a finite number of terms.

To illustrate the technology, let us look at the example of the leading Regge trajectory \eqref{leading} and its clone at $w=2\,$. The only possible operators that carry $2$ units of energy and act non--trivially on \eqref{leading}, which contains only oscillators of the type $\alpha_{-1}^\mu$, are $T^{11}$, $T^3{}_1$ and $(T^2{}_1)^2$ . The Ansatz for the clone thus takes the form
\begin{align}\label{ansatzlead}
    F_{w=2}=  \Big[ \beta_1 T^{11} + \beta_2 T^3{}_1+\beta_3(T^2{}_1)^2 \Big] \,F_{w=0}\,,
\end{align}
where $\beta_1\,$, $\beta_2$ and $\beta_3$ are a priori arbitrary coefficients. Notice that the relevant algebra in this particular case is $\mathfrak{sp}(6)\,$. Supplying the Virasoro constraints\footnote{The $L_0$ constraint \eqref{virfin} simply checks that the level of this clone is indeed $s+2$.} \eqref{virfin2} with the Ansatz \eqref{ansatzlead} and using the algebra \eqref{sp1}--\eqref{sp4}, one may write the solution in the following parametrization
\begin{align} \label{sol}
 \beta_2=- \beta_1 \displaystyle \frac{D+2 s-1}{s} \, , \quad \beta_3=\beta_1 \frac{D+2 s-1}{s(s-1)}\,,
\end{align}
where $D=26$. Consequently, the polynomial of the physical clone has an overall (unphysical) parameter $\beta_1$, while the rest of the coefficients depend on $D$ and a \textit{free} parameter, namely the spin $s$. The \textit{entire} trajectory of spin--$s$ states at depth $w=2$ is thus now known and tuning the spin to a given value immediately yields the values of the coefficients \eqref{sol}, hence the polynomial of the respective physical state. In other words, infinitely many states are excavated at the same cost, namely upon solving the Virasoro constraints \textit{only once}. Let us look at the lightest member--states of the clone. First, the cases $s=0,1$ are not allowed, as should be the case in accordance with the old methodologies, as there is no spin--$0$ or spin--$1$ at $w=2$ in table \ref{table:three}; in the parametrization of the solution as in \eqref{sol}, this is manifest via the divergence of the coefficients $\beta_2\,$ and $\beta_3$ for these spin values, but the result is the same in any parametrization: for example, for $s=0\,$, the polynomial collapses to the trivial function $F_{w=0}=1$, on which the operators $T^3{}_1$, $(T^2{}_1)^2$ act trivially, so the Ansatz contains only the $\beta_1$ term and solving the Virasoro constraints yields $\beta_1=0\,$, producing no physical polynomial. Indeed, as in table \ref{table:three}, the trajectory starts with a massive spin--$2$ state, whose physical polynomial takes, up to an overall coefficient, the form
\begin{align}
    F^{  \Yboxdim{4pt} \gyoung(;;)}_{w=2}=  \varepsilon_{\mu_1 \mu_2} \Big( 12\, \alpha_{-1} \cdot \alpha_{-1} \alpha_{-1}^{\mu_1} \alpha_{-1}^{\mu_2} -116\, \alpha_{-3}^{\mu_1} \alpha_{-1}^{\mu_2} +87\, \alpha_{-2}^{\mu_1} \alpha_{-2}^{\mu_2}  \Big)\,,
\end{align}
by substituting for $s=2$ in \eqref{ansatzlead}, \eqref{sol}. Interestingly, the multiplicity of a given trajectory (or state) is now given by the number of solutions to the Virasoro constraints for the coefficients of the Ansatz. In the simple example of the clone of the leading Regge at $w=2$, there is only one solution, hence one trajectory. Nontrivial multiplicity is, however, possible for $w>2$ and/or clones of more than $1$ row; for examples the reader is referred to the original work \cite{Markou:2023ffh}. As a last comment, let us emphasize that knowing whether a certain clone appears at a given depth and what its lightest member--state is is not a condition for the technology to work; rather, both are \textit{consequences} of the Virasoro constraints applied to the general Ansatz constructed with the technology at the desired depth.

\section{On the superstring, super quickly}

We begin with a brief review of the essential ingredients of the Ramond--Neveu--Schwarz (RNS) formulation of the (critical open) superstring, see \cite{Brink:1976sc,Friedan:1985ge} and, for example, the classic reference books \cite{Green:1987sp,Blumenhagen:2013fgp,Polchinski:1998rr}. In the oscillator language, a first difference with the (critical open) bosonic string is the appearance, additionally to the bosonic oscillators $\alpha_n^\mu$, of fermionic oscillators $b_{r}^\mu\,$,  $r \in \mathbb{Z}+\phi$, $\phi=\tfrac12,0$ (the meaning of the parameter $\phi$ will be made clearer in a bit), which are the Fourier modes of the field $\psi^\mu$\,,
\begin{align} \label{expansionf}
        \psi^\mu(z)  = \sum_{r \in \mathbb Z+\phi} b_r^\mu\,z^{-r-\frac12} \quad \Rightarrow \quad   b^\mu_{-r+\phi<0}&=\oint \frac{\md z}{2 \pi i}
    \frac{1}{z^{r+\nicefrac12-\phi}}\,\psi^\mu(z)\,,
\end{align}
$\psi^\mu$ being the worldsheet superpartner of the string field $X^\mu$\, that is also a spacetime vector as $X^\mu$. The fermionic oscillators can equivalently be denoted by $b_{r+\nicefrac12}^\mu$ and $b_{r}^\mu\,$,  $r \in \mathbb{Z}$, in the Neveu--Schwarz (NS) and Ramond (R) sectors, the two sectors of the open superstring spectrum, respectively: these Fourier modes are half--integer and integer respectively, due to the boundary conditions imposed on the field $\psi^\mu$ at the boundary of the open string worldsheet being periodic and antiperiodic respectively. They satisfy the algebra
\begin{align}\label{oscalg2}
    \{b^\mu_r , b^\nu_s\} = \delta_{r+s}\,\eta^{\mu\nu}\,, \quad r, s \in \mathbb{Z}+\phi\,,
\end{align}
where
\begin{align}
    \textrm{NS}\,: \quad\phi = \tfrac12 \,,\quad \textrm{R}\,: \quad \phi= 0 \,.
\end{align}
 The construction of the Fock spaces of states in the two sectors begins with the definition of the respective vacua via 
\begin{align}
    \alpha_m^\mu\,\ket{0,p}_{\textrm{NS}} &= 0 = b_r^\mu\,\ket{0,p}_{\textrm{NS}}\,,   \qquad \forall\, m,r>0 \\
    \alpha_m^\mu\,\ket{0,A,p}_{\textrm{R}} &= 0 = b_r^\mu\,\ket{0,A,p}_{\textrm{R}}\,,   \qquad \forall\, m,r>0 \,,
\end{align}
where the R vacuum carries a Dirac spinor index $A$ of $\mathfrak{so}(D-1,1)$. This is due to the presence of certain fermionic oscillators in the R sector, namely $b_0^\mu$, which carry no units of energy (like the bosonic $\alpha_0^\mu \sim p^\mu$), but satisfy a Clifford algebra $    \{b_0^\mu, b_0^\nu\} =\eta^{\mu\nu}$ (unlike $[\alpha_0^\mu,\alpha_0^\nu]=0$), because of which one may identify
\begin{align}\label{eq:gamma}
    b_0^\mu := \tfrac{1}{\sqrt{2}}\gamma^\mu\,,
\end{align}
where $\gamma^\mu$ are the $\gamma$ matrices in $D$ dimensions. This implies that the NS vacuum is a scalar but the R vacuum a spinor, and so all NS states are spacetime bosons and all R states spacetime fermions. A number operator for the fermionic oscillators can be defined via
\begin{align}
   N_r^F:=  b_{-r} \cdot b_r\,, \quad \forall\, r \in \mathbb{Z}+\phi>0\,.
\end{align}

The critical dimension is now $D=10$, in which physical superstring states $\ket{\textrm{phys}}$ are those functions of $\alpha_{-n}^\mu$ and $b_{-r}^\nu$ that pass the super--Virasoro constraints
\begin{align} \label{superVir}
    \big( L_0-\phi \big) |\textrm{phys} \rangle &= 0\,, \\ \label{conosc1}
    L_m |\textrm{phys} \rangle &= 0\,, \quad m>0\,, \\ \label{conosc2}
    G_r |\textrm{phys} \rangle &= 0\,, \quad   r  \geq \phi\,,
\end{align}
where $L_n$ and $G_r$, given by 
\begin{align}
    L_n &= \tfrac12 \sum_{m\in \mathbb{Z}}
    :\alpha_{-m} \cdot \alpha_{m+n}:
    +\, \tfrac12 \sum_{r\in \mathbb{Z}+\phi}
    \big(r+\tfrac{n}{2}\big) :b_{-r} \cdot b_{n+r}: \\
    G_r & =\sum_{m\in \mathbb{Z}} :\alpha_{-m} \cdot b_{m+r}:\,,
\end{align}
are the Fourier modes of the worldsheet energy--momentum tensor and its worldsheet superpartner, namely the supercurrent, and generate the super--Virasoro algebra. Because of the latter, sufficient are the following constraints (see also \cite{Basile:2024uxn}),
\begin{align}\label{suffsuperVir}
    \textrm{NS}\, &:\,
    \Big( L_0- \frac{1}{2} \Big)\ket{\textrm{phys}} = 0\,,
    \quad G_{\nicefrac12} \ket{\textrm{phys}} = 0\,,
    \quad G_{\nicefrac32} \ket{\textrm{phys}} = 0\,, \\
    \textrm{R}\, & :\, G_0 \ket{\textrm{phys}} = 0\,,
    \quad G_1 \ket{\textrm{phys}} = 0\,.
\end{align}
Defining the level $N$ as
\begin{align}
    N :=\sum_{m\in \mathbb{Z}} N_m^B +  \sum_{r \in \mathbb{Z}+\phi>0} r N_r^F =  \sum_{m\in \mathbb{Z}} \alpha_{-m} \cdot \alpha_m
    + \sum_{r \in \mathbb{Z}+\phi>0} r\,b_{-r} \cdot b_r\,,
\end{align}
which is half--integer and integer in the NS and R sectors respectively, the $L_0$ constraint can be rewritten as
\begin{align}
M^2=\frac{N-\phi}{\alpha'}\,.
\end{align}
The first few levels of the spectrum are given in table \ref{table:four} \cite{Hanany:2010da,Koh:1987hm,Feng:2010yx}, see also \cite{Basile:2024uxn}, in terms of the Young diagrams corresponding to the polarizations of the physical states in question. After the GSO projection, which essentially removes all integer--level valued levels of the NS sector (highlighted in gray in table \ref{table:four}) and half of the chiralities of all states in the R sector, the spectrum is rendered tachyon--free and is organized in multiplets of $\mathcal{N}=1$ spacetime supersymmetry at every level. All states are irreps of $\mathfrak{so}(D-1)$ except from the first level, that is now massless and contains the vector multiplet with components corresponding to irreps of $\mathfrak{so}(D-2)\,$. The leading Regge trajectories are highlighted in red in table \ref{table:three}: the one in the NS sector takes the form \cite{Schlotterer:2010kk}
\begin{align} \label{leadingNS}
\ket{\textrm{leading}}_{\textrm{NS}}:=\varepsilon_{\mu_1 \dots \mu_s} \, \alpha_{-1}^{\mu_1} \dots \alpha_{-1}^{\mu_{s-1}} b_{-\nicefrac12}^{\mu_s}  \ket{0;p}\,,
\end{align}
on which $G_{\nicefrac12}$ and $G_{\nicefrac32}$ impose transversality and tracelessness (the latter with respect to the indices contracted with $\alpha$'s and the index contracted with $b$) respectively, while its spacetime superpartner takes the form \cite{Schlotterer:2010kk}
\begin{align} \label{leadingR}
\ket{\textrm{leading}}_{\textrm{R}}:=\upsilon_{\mu_1 \dots \mu_{s-1}}^a \, \alpha_{-1}^{\mu_1} \dots \alpha_{-1}^{\mu_{s-1}} \ket{0;a;p}  + \rho_{\dot a \, \mu_1 \dots \mu_{s-1}} \, \alpha_{-1}^{\mu_1} \dots \alpha_{-1}^{\mu_{s-2}} b_{-1}^{\mu_{s-1}} \ket{0;\dot a;p} \,,
\end{align}
where $a$, $\dot a$ are Weyl spinor indices corresponding to the two possible chiralities of the vacuum before the GSO, such that the two terms in the expression \eqref{leadingR} have the same chirality, thus surviving the GSO. The constraint $G_0$ checks then the massive Dirac equation on $\upsilon_{\mu_1 \dots \mu_{s-1}}^a$, $\rho_{\dot a \, \mu_1 \dots \mu_{s-1}}$, while $G_1$ imposes transversality and $\gamma$--tracelessness. It is worth noting that tracelessness with respect to all indices contracted with $\alpha$ oscillators is a consequence of $L_2$ in both sectors, which, due to the super--Virasoro algebra, is itself a consequence of $\{G_{\nicefrac12}, G_{\nicefrac32} \}$ in the NS and of $\{G_1,G_1\}$ in the R sector respectively.

\begin{table}
\centering 
\renewcommand{\arraystretch}{1.5}
  \begin{tabular}{ c || l | l  }
   $M^2$ & NS & R  \\ \hline \hline
   $-\tfrac12$ & $\textcolor{gray}{\bullet}$ &  \\
   $0$ & $\LeadingB{;}^{\textcolor{red}{so(D-2)}}$
   & $\textcolor{red}{\bullet^{so(D-2)}_{\nicefrac12}}$ \\
   $\tfrac12$ & $\NoGSO{;,;}$ &  \\
   $1$ & $\LeadingB{;;} \oplus\, \gyoung(;,;,;)$
   & $\LeadingF{;}$  \\
   $\tfrac32$ & ${\color{gray}\bullet}\,
   \oplus\,  \NoGSO{;;}\, \oplus\, \NoGSO{;;,;} \oplus\, \NoGSO{;,;,;,;}\, $ & \\
   $2$ & $\LeadingB{;;;}\, \oplus\, \gyoung(;)\, 
   \oplus\, \gyoung(;;,;)\, \oplus\, \gyoung(;,;)\,
   \oplus\, \gyoung(;;,;,;)\, \oplus\, \gyoung(;,;,;,;,;)$ 
   & $\LeadingF{;;}\, \oplus\, \gyoung(;)_{\nicefrac12}\, 
   \oplus\, \bullet_{\nicefrac12}
   \oplus\, \gyoung(;,;)_{\nicefrac12}$
  \end{tabular}
\renewcommand{\arraystretch}{1}
\caption{Open superstring, physical content of the first few levels in terms of the respective Young diagrams. }
\label{table:four}
\end{table}

Now we are ready to review the original work \cite{Basile:2024uxn}. Before we turn to the construction of general superstring states, let us consider how states with polynomials of fermionic oscillators \textit{only} can be built, namely the ways in which the oscillators $b_{-r+\phi}^\mu$ ($r \in \mathbb{Z}$) may contract the indices corresponding to the boxes of a given Young diagram.  Due to the anticommuting nature of the $b_{-r}^\mu$, fermionic oscillators that carry the same amount of energy cannot contract boxes of the same row (in the symmetric base), so the antisymmetric base is favored, in which a general $\mathfrak{gl}$ tensor or the respective Young diagram may be denoted by
$\varepsilon_{\mu[h_1],\, \lambda [h_2], \dots,\, \nu[h_L]}\,$, where the notation $\mu[h_1]:=\mu_1 \dots \mu_{h_1}$ stands for $h_1$ \textit{antisymmetrized} indices. Each group of antisymmetrized indices corresponds to a different column, different groups are separated by commas and the height $h_i$ (namely the number of boxes in the $i$--th column) cannot be increasing as one moves along a given diagram to the right. It should be noted that either of the two bases are possible to choose, but for now we choose the antisymmetric one purely for convenience. Consequently, focusing on the polynomials that also keep the level as low as possible, in analogy with the construction of the $w=0$ trajectories of the bosonic string, the boxes of the $i$--th \textit{column} have to be contracted with $b_{-i+\phi}^\mu\,$, namely
\begin{align} \label{simplestNS}
    F_{\textrm{simpl}}^{\textrm{NS}} = \varepsilon_{\mu[h_1],\, \lambda [h_2], \dots,\, \nu[h_L]} \, b_{-\nicefrac12}^{\mu_1} \dots b_{-\nicefrac12}^{\mu_{h_1}}  b_{-\nicefrac32}^{\lambda_1} \dots b_{-\nicefrac32}^{\lambda_{h_2}} \dots b_{-L+\nicefrac12}^{\nu_1} \dots b_{-L+\nicefrac12}^{\nu_{h_L}} 
\end{align}
at level
\begin{align}
     N_{\text{min}}^{\textrm{NS}}=\sum_{i=1}^L h_i \,\big(i-\tfrac12\big) 
\end{align}
and
\begin{align} \label{simplestR}
    F_{\textrm{simpl}}^{\textrm{R}} = \varepsilon_{\mu[h_1],\, \lambda [h_2], \dots,\, \nu[h_L]} \, b_{-1}^{\mu_1} \dots b_{-1}^{\mu_{h_1}} b_{-2}^{\lambda_1} \dots b_{-2}^{\lambda_{h_2}}  \dots b_{-L}^{\nu_1} \dots b_{-L}^{\nu_{h_L}} 
\end{align}
at level
\begin{align}
     N_{\text{min}}^{\textrm{R}}=\sum_{i=1}^L h_i \,i\,.
\end{align}
Let us note that only one class of physical superstring states has polynomials built solely out of fermionic oscillators: these are the (first apparitions of) totally antisymmetric tensors.

Next, as in the bosonic case, let us construct all fermionic oscillator bilinears that are also spacetime scalars in the two sectors:
\begin{align} \label{opsNS}
  \textrm{NS}\,:\quad  M^r{}_s   := b_{-r+\nicefrac12}\cdot b_{s-\nicefrac12}  \, , \quad  M_{rs} :=  b_{r-\nicefrac12} \cdot b_{s-\nicefrac12}\, , \quad M^{rs}:= b_{-r+\nicefrac12}\cdot b_{-s+\nicefrac12} 
  \end{align}
and
\begin{align}   \label{opsR}
\begin{aligned}
   \textrm{R}\,:\quad  M^r{}_s &:= b_{-r}\cdot b_s  \, , \quad  M_{rs} :=  b_r \cdot b_s\, , \quad M^{rs}:= b_{-r}\cdot b_{-s}\,,\\ 
    \qquad M_r  & := b_0 \cdot b_r \,, \quad  M^r:=b_0\cdot b_{-r}\,,
\end{aligned}
\end{align}
where $r,s=1,2,\dots\,M$, with $M$ being an integer, and the prefactors differ from the bosonic case purely for convenience. The operators $M^r{}_s $, $M_{rs}$ and $M^{rs}$ carry $r-s$, $-r-s+2\phi$ and $r+s-2\phi$ units of energy respectively, while the operators $M_r$ and $M^r$ carry $-r$ and $+r$ units respectively. Employing the oscillator algebra \eqref{oscalg2}, it can be shown that (after the redefinition $M^r{}_s \rightarrow M^r{}_s + \tfrac{D}{2}  \delta^r_s$)  the $M$ operators \eqref{opsNS} and \eqref{opsR} generate the orthogonal algebras $\mathfrak{o}(2M)$ and $\mathfrak{o}(2M+1)$ in the NS and R sectors respectively. It is then possible to split their generators into lowering and raising operators, for example with the convenient choice
\begin{align}
    \textrm{lowering}&:\quad M_{rs}\,,\quad M^{r<s}{}_{s}\,,\quad M_r \\
 \textrm{raising}&:\quad M^{rs}\,,\quad M^{r>s}{}_{s}\,, \quad M^r\,,
\end{align}
since, as in the bosonic case, the raising and lowering operators now add and remove units of energy respectively. The lowest weight states $ \widetilde{F}$ are thus defined via 
    \begin{align}
 M_{rs} \widetilde{F}=0\,,\quad M^{r<s}{}_{s} \widetilde{F}=0  \,,\quad    M^{r}{}_{r} \widetilde{F}=h_r \,\widetilde{F}\,,\quad  M_r  \widetilde{F}=0\,.
\end{align}
Of course, the last condition involving the operator $M_r$ only appears in the R sector, where it imposes $\gamma$--tracelessness due to \eqref{eq:gamma}. Which are the lowest weight states of the orthogonal algebra? It is easy to see that these are the states filled as in \eqref{simplestNS} and \eqref{simplestR} in the two sectors. For example, on the hook (recall that we are in the antisymmetric base)
\begin{align} \label{hookzeroF}
    F_{\textrm{simpl}}^{  \Yboxdim{4pt} \gyoung(;;,;)} = \tilde{\varepsilon}_{\mu_1 \mu_2,\, \lambda } \, b_{-1}^{\mu_1}  b_{-1}^{\mu_{2}}  b_{-2}^{\lambda}\,,
\end{align}
we have that
\begin{align}
    M^1{}_2  F_{\textrm{simpl}}^{  \Yboxdim{4pt} \gyoung(;;,;)}&= 0 \quad  \Leftrightarrow \quad \tilde{\varepsilon}_{[\mu_1 \mu_2\lambda]}  =0 \,.
\end{align}

However, the vast majority of superstring states are functions involving both bosonic and fermionic oscillators. At $w=0$, we have learned how to glue rows together using the lowest weight conditions of the symplectic algebra, as well as how to glue columns together using the lowest weight conditions of the orthogonal algebra, but \textit{how are rows glued with columns} to form $w=0$ superstring trajectories? For that, let us look at the type of oscillator bilinears that are also spacetime scalars that we have not yet considered: those involving one bosonic and one fermionic oscillator, namely
\begin{align} \label{qNS}
\begin{aligned}
       \textrm{NS}\,:\quad     Q^{nr} &:= \tfrac1n\,\alpha_{-n} \cdot b_{-r+\frac12}\,,
    \quad 
    Q^n{}_r := \tfrac1n\,\alpha_{-n} \cdot b_{r-\frac12}\,, \\
    Q_n{}^r & := b_{-r+\frac12} \cdot \alpha_n\,,
    \quad 
    Q_{nr} := \alpha_n \cdot b_{r-\frac12}
    \end{aligned}
\end{align}
and
\begin{align}\label{qR}
\begin{aligned}
       \textrm{R}\,:\quad     Q^{nr} &:= \tfrac1n\,\alpha_{-n} \cdot b_{-r}\,,
    \quad 
    Q^n{}_r := \tfrac1n\,\alpha_{-n} \cdot b_{r}\,, \\
    Q_n{}^r & := b_{-r} \cdot \alpha_n\,,
    \quad 
    Q_{nr} := \alpha_n \cdot b_{r} \\
   Q^m &:= \tfrac1{m}\,b_0\cdot \alpha_{-m}\,,
    \quad 
    Q_m :=  b_0\cdot \alpha_m\,,
    \end{aligned}
\end{align}
where the modes of the bosonic and fermionic oscillators run from $1$ to $N$ and to $M$ respectively. The operators $Q^{nr}$, $Q^n{}_r$, $Q_n{}^r$ and $Q_{nr}$ carry $n+r-\phi$, $n-r+\phi$, $-n+r-\phi$ and $-n-r+\phi$ units of energy respectively, while the operators $Q_m$ and $Q^m$ carry $-m$ and $m$ units respectively. Together with the $T$ operators \eqref{opT} and $M$ operators \eqref{opsNS}, \eqref{opsR} (and after absorbing the $D$ contribution in the definition of the operators as previously), these form the algebras $\mathfrak{osp}(2M|2N)$ and $\mathfrak{osp}(2M+1|2N)$ in the NS and R sectors respectively, of which we quote a few commutators for illustration purposes:
    \begin{align}
    \begin{aligned}
     \{Q^m{}_r, Q^{ns}\} & = \delta^s_r\,T^{mn}\,,
        & \{Q_m{}^r, Q^{ns}\} & = \delta^n_m\,M^{rs}\,,\\
        \{Q^m{}_r, Q_{ns}\} & = -\delta^m_n\,M_{rs}\,,
        & \{Q_m{}^r, Q_{ns}\} & = \delta^r_s\,T_{mn}\,,\\
        \{Q^m{}_r, Q_n{}^s\} & = \delta_r^s\,T^m{}_n
        + \delta^m_n\,M^s{}_r\,,
        & \{Q^{mr}, Q_{ns}\} & = \delta^r_s\,T^m{}_n
        - \delta^m_n\,M^r{}_s\,,
            \end{aligned}
    \end{align}
while the rest can be found in \cite{Basile:2024uxn}. The next natural step is to split the generators $Q$ into raising and lowering operators. Similarly to the reason behind the choices for the purely bosonic and purely fermionic cases, we may identify
\begin{align} \label{choice1}
    \textrm{lowering}&:\quad Q_{nr}\,,\quad Q^{n< r}{}_{r}\,,\quad Q_{n\ge r}{}^r\,,\quad Q_r \\ \label{choice2}
 \textrm{raising}&:\quad Q^{nr}\,,\quad Q^{n\ge r}{}_{r}\,, \quad Q_{n<r}{}^r\,, \quad Q^r
\end{align}
in the NS sector, since the former and latter types of operators remove and add units of energy respectively in the sector in question. With this choice, it is the lowering operators of the orthosymplectic algebra that appear in the super--Virasoro constraints in the NS sector. In particular, using the definitions \eqref{qNS}, \eqref{qR} and the identification \eqref{eq:gamma}, the sufficient super--Virasoro constraints take the following form in the transverse subspace simplification:
\begin{align}
    \label{VirTNS1}
    \Big(L_0-\tfrac12 \Big) F_{\textrm{NS}} 
    & = \Big[  N_{\textrm{NS}} 
    +\tfrac12(p^2 -1)  \Big] \, F_{\textrm{NS}} = 0\,, \\ 
    \label{VirTNS2}
    G_{\nicefrac12} F_{\textrm{NS}}
    & = \sum_{n\geq1}\, \Big[ n\,  Q^n{}_{n+1}
    + Q_{n}{}^n \Big] F_{\textrm{NS}} = 0\,,  \\ \label{VirTNS3}
    G_{\nicefrac32} F_{\textrm{NS}} & = \Big[ Q_{11}
    + \sum_{k\geq1} \Big( k\,Q^k{}_{k+2}
    + Q_{k+1}{}^k \Big) \Big]F_{\textrm{NS}} = 0\,,
\end{align}
where the level is given by
\begin{align}
        N_{\textrm{NS}}  F_{\textrm{NS}}
    = \sum_{n\geq1} \Big[ n \, T^n{}_n
    + \big( n-\tfrac12 \big) M^n{}_n \Big] F_{\textrm{NS}}\,.
\end{align}

The choice \eqref{choice1}--\eqref{choice2} also tells us how rows are to be glued to columns at $w=0$ in the NS sector. More specifically, let us consider a diagram of $K$ rows contracted with $\alpha$'s and a diagram of $L$ columns contracted with $b$'s as in $w=0$ (both in the same base) and imagine that the rows and columns are somehow glued together to form a new Young diagram with polynomial $\widetilde F$. The two diagrams are lowest weight states of $\mathfrak{sp}$ and $\mathfrak{o}$ respectively and now let us impose that their ``fusion'' be annihilated by all lowering operators of $\mathfrak{osp}$. The lowest weight conditions of the $\mathfrak{osp}(2M|2N)$ according to \eqref{choice1} include
\begin{align} \label{nest}
    Q_n{}^n \widetilde{F} = 0\,,
\end{align}
which, since the operators $Q_n{}^n $ replace $\alpha_{-n}$ with $b_{-n+\nicefrac12}\,$, implies that completely antisymmetrizing the row $n$ with one box from the column $n$ be zero and so that the top of $n$--th columnn has to be glued to the left of the $n$--th row, yielding a ``nested--hook--like'' structure. This is now the recipe for all $w=0$ trajectories in the NS sector. For example, the $3$--row trajectory at $w=0$ has a polynomial as indicated below:
\begin{align}
    \Yboxdim{24pt}\scriptscriptstyle
    \gyoung(\boneh;\alphaone_8{\cdots};\alphaone,%
    \boneh;\bthreeh;\alphatwo_6{\cdots};\alphatwo,%
    \boneh;\bthreeh;\bfiveh;\alphathree_3{\cdots};\alphathree)
    \end{align}
Obtaining trajectories at $w>0$ is then straightforward by dressing the $w=0$ trajectories, namely the $\mathfrak{osp}(2M|2N)$ lowest weight states, with Ans\"atze constructed using the raising operators of the $\mathfrak{osp}(2M|2N)$ algebra and can be effectuated equally well before and after the GSO projection. The reader is referred to \cite{Basile:2024uxn} for examples of trajectories and clones.

Turning to the R sector, a first observation is that the operators $Q^n{}_n$ and $Q_n{}^n$ now carry \textit{zero} units of energy. This means that the condition \eqref{nest} is no longer necessarily satisfied for $w=0$ trajectories, which implies that the top of the $n$--th column may be glued to the left of \textit{or} under the left end box of the $n$--th row at $w=0$. In other words, the \textit{diagonal} of a given diagram may be filled by either $\alpha_{-n}$ or $b_{-n}$, since they carry the same units of energy. For the three--row trajectory for example, this means that we know how to contract its boxes up to the ambiguity along the diagonal
\begin{align}
        \Yboxdim{20pt}\scriptscriptstyle
    \gyoung(;;\alphaone_9{\cdots};\alphaone,%
    \bone;;\alphatwo_5{\cdots};\alphatwo,%
    \bone;\btwo;;\alphathree_2{\cdots};\alphathree)
    {\textstyle}
\end{align}
which yields $2^K$ different configurations, $K$ being the diagonal's length. Since, for any $K$, one  of the possible configurations is given by contracting all boxes of the diagonal with $b$ oscillators, as in the $w=0$ trajectories in the NS sector, while the operators $Q^n{}_n$ replace the oscillator $b_{-n}$ with $\alpha_{-n}$ without changing the level, a general Ansatz for the $w=0$ trajectories in the R sector can take the form
\begin{align}\label{candidate}
   F^{w=0}_R
    := \bigg( \beta_0+\sum_{n=1}^{K} \tfrac1{n!}\,\beta_{i_1 \dots i_n}\,
    Q^{i_1}{}_{i_1} \dots Q^{i_n}{}_{i_n} \bigg) \widetilde{F}_R\,,
\end{align}
where the $\beta$'s are a priori arbitrary parameters and $\widetilde{F}_R$ is the polynomial for an a priori arbitrary polarization tensor--spinor of $n$ columns contracted with $\alpha_{-i}$'s and $b_{-i}$'s precisely like the $w=0$ trajectories in the NS sector, namely as in the nested--hook--like structure, but with \textit{integer} instead of half--integer $b$'s. In other words, the $ \widetilde{F}_R$ \textit{are} lowest--weight states of $\mathfrak{osp}(2M+1|N)$, namely they are annihilated by all its lowering operators by construction, but $ F^{w=0}_R$ are \textit{not}. Note that the order of the operators $Q$ in the Ansatz \eqref{candidate} carries no meaning, as they all anticommute. 

Now, the super--Virasoro constraints take the form
\begin{align}
    \label{VirTR1}
    G_0 F_{\textrm{R}} & = \Big[ \tfrac1{\sqrt{2}}\,\slashed{p}
    + \sum_{k\geq1} \Big(k\,Q^k{}_{k} + Q_k{}^k\Big) \Big] 
    F_{\textrm{R}} = 0 \,,\\  \label{VirTR2}
    G_1 F_{\textrm{R}} & = \Big[ Q_1 + \sum_{k\geq1}\Big(k\,Q^k{}_{k+1} 
    + Q_{k+1}{}^k\Big) \Big] F_{\textrm{R}} = 0\,,
\end{align}
with the level given by
\begin{equation}
    N_{\textrm{R}} F_{\textrm{R}}
    = \sum_{n\geq1} n \, \Big( T^n{}_n
    +  M^n{}_n \Big)F_{\textrm{R}} \,.
\end{equation}
It can then be shown that, imposing the $G_0$ constraint on the Ansatz \eqref{candidate}, one may solve for the coefficients $\beta$ and obtain a \textit{multiplicity} of $2^{K-1}$ solutions, while the $G_1$ constraint checks the $\gamma$--tracelessness of the tensor--spinor of $\widetilde{F}_R$. The appearance of nontrivial multiplicity even at $w=0$ is a feature of the R sector, while in the NS sector, as well as in the purely bosonic string, it is only possible for $w>0$, as we have seen. This result is not only in accordance with the multiplicities obtained from the partition function of \cite{Hanany:2010da}, but can also be intuitively understood in the context of spacetime supersymmetry, in the sense that several fermions of the same spin may be needed to close the supersymmetry algebra on two distinct \textit{massive} supermultiplets at the same level. For example, the first apparition of the ``window'' diagram $\gyoung(;;,;;)_{\nicefrac12}$, where the index $\nicefrac12$ signifies that the corresponding polarization is a tensor--spinor, finds itself at level $5$ with multiplicity $2^{2-1}=2$, based on the technology just reviewed. Indeed, as can be deduced from the partition function, the two windows belong to two distinct supermultiplets. Obtaining $w>0$ trajectories can now proceed as in the previous cases, by employing the raising operators of $\mathfrak{osp}(2M+1|2N)$ \textit{including} the zero--energy operators $Q^n{}_n$. For more details and examples of trajectories and clones, the reader is referred to \cite{Basile:2024uxn}.

\section{Conclusions}

Let us summarize what we have learned so far. String spectra consist of infinite towers of physical states, which, at least in principle, can be constructed with the traditional methodologies on a level--by--level basis. A generic string state is a function of creation oscillators $\alpha_{-n}^\mu\,$ acting on the vacuum and contracted with an appropriate polarization tensor, which encodes the state's spacetime symmetries as an irreducible representation of $\mathfrak{so}(d-1)$ (and  $\mathfrak{so}(d-2)$  for the very few massless states) and which can be represented by a Young diagram. A given Young diagram may be contracted with the oscillators in infinitely many different ways, which embed it within infinitely many distinct physical states in the spectrum, albeit at different mass levels (up to multiplicity). The new technology begins then by defining the ``depth'' $w$ of a given state as the difference between its level and the level of the ``first'' apparition of its Young diagram in a physical state, namely the lowest possible level at which the same diagram can be embedded in the spectrum. Consequently, the spectrum may be split into two parts, one with all possible trajectories at $w=0$, and another with their infinitely many clones, in terms of their respective Young diagrams, at $w>0$. The former have simple and known polynomials, while the latter more complicated ones, which are precisely what the new technology excavates.

In the open bosonic string, the key observation is then that the Virasoro constraints, which impose physicality of states, are in fact linear combinations of the lowering operators of a larger algebra, that is a symplectic algebra $\mathfrak{sp}$, which acts on the modes $n$ of the oscillators $\alpha_{n}^\mu\,$. It is then straightforward to see that the $w=0$ trajectories of the spectrum are precisely the lowest weight states of the $\mathfrak{sp}$ algebra. Moreover, this algebra commutes with the spacetime Lorentz algebra, (of the little group) of which all physical string states are irreducible representations. Consequently, it is possible to employ Howe duality, which relates the irreps of the two algebras via a bijection. This means that a given $\mathfrak{so}$ irrep corresponds to a certain $\mathfrak{sp}$ irrep and the first apparition of the former (within a physical state) to the lowest weight state of the respective $\mathfrak{sp}$ irrep, so that every other apparition can be be reached by means of the raising operators of the $\mathfrak{sp}$, the gearwheel of the technology! The technology works namely as follows. To construct the clone, at a certain depth, of a given $w=0$ trajectory, dress the latter by a polynomial constructed out of all possible raising $\mathfrak{sp}$ operators that carries as many units of energy as the depth one aims to reach. Then, apply the Virasoro constraints to the Ansatz and solve for its a priori arbitrary coefficients. The solution is the physical polynomial for the entire clone, with the coefficients now being functions of the lengths of the rows of the Young diagram in question. By solving the Virasoro constraints only once, the entire clone is excavated and all infinitely many  member--states are found at no extra cost! The technology thus comes with a huge improvement in the \textit{efficiency} of constructing string states.

In the superstring, the construction in the Neveu--Schwarz sector is closely resembling that of the bosonic string, albeit with more ingredients. Additionally to the bosonic oscillators there appear fermionic ones which carry half--integer units of energy, and the algebra Howe--dual to the spacetime Lorentz is now a larger orthosymplectic algebra $\mathfrak{osp}$. Its lowest weight conditions provide a recipe for constructing trajectories at $w=0$: rows and columns are contracted with bosonic and fermionic oscillators respectively, forming a ``\textit{nested--hook--like}'' configuration. Trajectories at $w>0$ can again be constructed by means of the raising operators of the $\mathfrak{osp}$, so the Ans\"atze are slightly larger compared to the bosonic case. In the Ramond sector however, the lowest weight states of the $\mathfrak{osp}$ are not sufficient to construct the $w=0$ trajectories, like in the Neveu--Schwarz sector. This difference can be traced back to the presence of two distinct types of oscillators that carry the same amount of units of energy, since now the fermionic oscillators carry integer units of energy, and which is responsible for the appearance of certain generators of the $\mathfrak{osp}$ that carry no units of energy. The latter can then be employed to write general Ans\"atze for the $w=0$ trajectories in the Ramond sector, supplying with which the Virasoro constraints yields a nontrivial \textit{multiplicity} even at $w=0$, unlike the Neveu--Schwarz sector or the bosonic string. Dressing the solutions with the raising (including the zero--energy) $\mathfrak{osp}$ operators and solving the Virasoro constraints yields again physical clones, as in the previous cases. 

Why are we interested in constructing string states and trajectories? Firstly, let us highlight that it is the polynomial of a given state that encodes the information on how the state \textit{interacts}. More specifically, string scattering amplitudes are essentially the correlation functions of the polynomials, in the equivalent CFT language, of the external states (integrated over the insertion points of the states on the string worldsheet). These are computed by performing operator contractions via Wick's theorem, so two physical string states with the same Young diagram but different polynomials a priori have very different scattering amplitudes. For example, curiously, the $3$--point amplitudes of certain among the lightest closed string massive spin--2 states can even reproduce \cite{Lust:2023sfk} the (on--shell) cubic vertices of a \textit{field} theory, namely of the extension of dRGT massive gravity \cite{deRham:2010kj} to ghost--free bimetric theory \cite{Hassan:2011zd}, which in a maximally symmetric background propagates the graviton and a massive spin--2 state \cite{Hassan:2011ea}. However, the same is not true for the lightest open string massive spin--2 states \cite{Lust:2021jps}, while at the moment there is no reason to (not) expect that it is true for the rest of the infinitely many massive spin--2 states of string spectra. Turning to the scattering of entire trajectories deeper in the string spectrum, a generalized\footnote{With ``generalized'' here we refer to the correlation function of the exponentiated $w>0$ polynomials of the external legs, the action of suitable differential operators per trajectory on which gives the standard form of the respective polynomials.} $n$--point Koba--Nielsen factor for any bosonic string trajectories in the external legs is given in \cite{Markou:2023ffh}, together with formulae for $3$--point amplitudes of trajectories at $w=0$ and at $w>0$ as well as several examples. 

In the Neveu--Schwarz sector of the superstring, the state--operator correspondence works in a similar way as in the bosonic string, since the expansion \eqref{expansionf} yields
\begin{align}
     \ket{p;0}_\textrm{NS} = \lim_{z\to0} e^{ip \cdot X(z)} \ket0 \,,\quad   b^\mu_{-r+\frac12} \ket{0}
    \, \leftrightarrow \,
    \frac{1}{(r-1)!}\,\partial^{r-1} \psi^\mu(0)\,,
\end{align}
so it suffices to replace bosonic and fermionic oscillators with the respective descendants to obtain vertex operators\footnote{Picture changing is also possible using the technology, see \cite{Basile:2024uxn} for a formula for a generic NS state.}. However, the Ramond sector is slightly more involved, since its vacuum is a fermion: it is mapped to the \textit{spin--field} $S_A$, which is a Dirac spinor of $\mathfrak{so}(D-1,1)$ and which interacts with $\psi^\mu$, acting on the NS vacuum
\begin{align}
        \ket{p; A;0}_\textrm{R} =\lim_{z\to0} S_A(z)  \ket{p;0}_\textrm{NS} =  \lim_{z\to0} S_A(z) \,e^{ip\cdot X}(z)\ket0\,.
\end{align}
The expansion \eqref{expansionf} hence yields
\begin{align}
    b_{-r}^\mu  \ket{p; A;0}_{\textrm{R}}
    = \lim_{w\rightarrow 0}  \oint \frac{\md z}{2 \pi i}
    \frac{1}{z^{r+\nicefrac12}} \psi^\mu(z) S_A(w) \ket{p;0} \,,
\end{align}
so there is no \textit{explicit} dictionary mapping any fermionic oscillator to $\psi^\mu$ and its descendants;  rather, the dictionary has to be computed for the desired oscillator every time a new one is excited. Consequently, finding the operator version of a string state (and its amplitudes) in the Ramond sector involves a few more steps than in the Neveu--Schwarz, but is of course still possible if needed.

Let us mention other possible applications of the technology. First, in the context of \textit{field} theory, higher--spins have been shown to capture properties of what has come to be known as \textit{Kerr black hole} scattering amplitudes, see indicatively \cite{Guevara:2018wpp,Maybee:2019jus,Cangemi:2022bew}, while in the case of \textit{string} theory, the leading Regge trajectory in the NS sector was found not to share this property. Since the reason behind this mismatch is related to the spin multipole coefficients coming from leading Regge amplitudes being spin--dependent, it would be interesting to explore whether there exist subleading trajectories or other combinations that respect the \textit{spin--universality} of black hole amplitudes\footnote{The author thanks Lucile Cangemi and Paolo Pichini for interesting discussions on this point.}. In a different direction, since closed string scattering off D--branes models aspects of \textit{Hawking radiation} emission and absorption and only massless or some of the lightest open strings and the leading Regge have been cast as external states of such amplitudes, see for example \cite{Callan:1996dv,Dhar:1996vu,Das:1996wn,Maldacena:1996ix,Hashimoto:1996bf,Bianchi:2017sds,Bianchi:2018kzy} (and also \cite{Lust:2021jps}), it would be exciting to probe deeper trajectories as excitations of D--branes. Other than that, the decay rates of subleading trajectories could be of interest to the study of the possible \textit{chaotic traits}, see for example \cite{Gross:2021gsj,Bianchi:2022mhs,Pesando:2024lqa}, of the decays of highly excited strings. Finally, developing the technology in the phenomenologically more interesting cases of $4$ spacetime dimensions and \textit{curved} backgrounds is a natural next direction, bearing in mind that a lot is known for certain tensionless ($\alpha' \rightarrow \infty$) string spectra, in which all higher spins are massless, via the AdS--CFT correspondence \cite{Bianchi:2003wx,Beisert:2003te,Beisert:2004di,Gaberdiel:2018rqv} and the related longstanding speculation of the community that string theory may correspond to a broken phase of higher spin gravity, see, for example, \cite{Sagnotti:2010at,Sundborg:2000wp,Taronna:2011kt}. Let us finally highlight that solving the Virasoro constraints for \textit{arbitrary} depth $w$ is a natural next challenge to pursue by means of the technology, which would in principle enable us to reach the \textit{entire} string spectrum. After all, it is the infinitely many higher spins of the string spectrum that are precisely at the origin of the good UV behavior of string amplitudes, one of the most attractive features of string theory. The subleading Regge trajectories may hence be holding the key to understanding aspects of the underlying structure of string theory per se. 

\paragraph{Acknowledgements.}The present proceedings material is meant to be a pedagogical and brief review of the original published work of the author with Evgeny Skvortsov for the open bosonic string \cite{Markou:2023ffh} and with Thomas Basile for the open superstring \cite{Basile:2024uxn}; the author thanks the organizers of the Corfu Summer Institute 2024 for the kind invitation to give a talk. The author is very grateful to Evgeny Skvortsov, from whom she learned a great deal, for enlightening discussions of formative effect; she further thanks Thomas Basile for very beneficial interactions and collaboration, as well as for valuable comments on a draft of this presentation. In addition, the author extends her sincere thanks to Augusto Sagnotti for illuminating and most helpful discussions. It is also a pleasure to acknowledge scientific interactions with Dionysios Anninos, Massimo Bianchi, Paolo Di Vecchia, Amihay Hanany, Henrik Johansson, Euihun Joung, Renann Lipinski Jusinskas, Oliver Schlotterer, Stephan Stieberger and Arkady Tseytlin, which have improved this presentation. The author is supported by a fellowship of the Scuola Normale Superiore and by INFN (I.S. GSS-Pi).


\begin{thebibliography}{99}

%\cite{Markou:2023ffh}
\bibitem{Markou:2023ffh}
C.~Markou and E.~Skvortsov,
``An excursion into the string spectrum,''
JHEP \textbf{12} (2023), 055
%doi:10.1007/JHEP12(2023)055
[arXiv:2309.15988 [hep-th]].

%\cite{Basile:2024uxn}
\bibitem{Basile:2024uxn}
T.~Basile and C.~Markou,
``On the deep superstring spectrum,''
JHEP \textbf{07} (2024), 184
%doi:10.1007/JHEP07(2024)184
[arXiv:2405.18467 [hep-th]].

%\cite{Kato:1982im}
\bibitem{Kato:1982im}
M.~Kato and K.~Ogawa,
``Covariant Quantization of String Based on BRS Invariance,''
Nucl. Phys. B \textbf{212} (1983), 443-460
%doi:10.1016/0550-3213(83)90680-6


%\cite{Hwang:1982mc}
\bibitem{Hwang:1982mc}
S.~Hwang,
``Covariant Quantization of the String in Dimensions D \ensuremath{<}= 26 Using a BRS Formulation,''
Phys. Rev. D \textbf{28} (1983), 2614
%doi:10.1103/PhysRevD.28.2614

%\cite{Manes:1988gz}
\bibitem{Manes:1988gz}
J.~L.~Manes and M.~A.~H.~Vozmediano,
``A Simple Construction of String Vertex Operators,''
Nucl. Phys. B \textbf{326} (1989), 271-284
%doi:10.1016/0550-3213(89)90444-6

%\cite{Green:1987sp}
\bibitem{Green:1987sp}
M.~B.~Green, J.~H.~Schwarz and E.~Witten,
``SUPERSTRING THEORY. VOL. 1: INTRODUCTION,''
1988 
%,ISBN 978-0-521-35752-4

%\cite{Polchinski:1998rq}
\bibitem{Polchinski:1998rq}
J.~Polchinski,
``String theory. Vol. 1: An introduction to the bosonic string,''
Cambridge University Press, 2007
%,ISBN 978-0-511-25227-3, 978-0-521-67227-6, 978-0-521-63303-1
%doi:10.1017/CBO9780511816079

%\cite{Blumenhagen:2013fgp}
\bibitem{Blumenhagen:2013fgp}
R.~Blumenhagen, D.~L\"ust and S.~Theisen,
``Basic concepts of string theory,''
Springer, 2013
%,ISBN 978-3-642-29496-9
%doi:10.1007/978-3-642-29497-6

%\cite{DelGiudice:1970dr}
\bibitem{DelGiudice:1970dr}
E.~Del Giudice and P.~Di Vecchia,
``Characterization of the physical states in dual-resonance models,''
Nuovo Cim. A \textbf{70} (1970), 579-591
%doi:10.1007/BF02734495

%\cite{Fubini:1971ce}
\bibitem{Fubini:1971ce}
S.~Fubini and G.~Veneziano,
``Algebraic treatment of subsidiary conditions in dual resonance models,''
Annals Phys. \textbf{63} (1971), 12-27
%doi:10.1016/0003-4916(71)90295-8

%\cite{Sasaki:1985py}
\bibitem{Sasaki:1985py}
R.~Sasaki and I.~Yamanaka,
``Vertex Operators for a Bosonic String,''
Phys. Lett. B \textbf{165} (1985), 283-288
%doi:10.1016/0370-2693(85)91231-6

%\cite{Friedan:1985ge}
\bibitem{Friedan:1985ge}
D.~Friedan, E.~J.~Martinec and S.~H.~Shenker,
``Conformal invariance, supersymmetry and string theory,''
Nucl. Phys. B \textbf{271} (1986), 93-165
%doi:10.1016/S0550-3213(86)80006-2

%\cite{Sagnotti:2010at}
\bibitem{Sagnotti:2010at}
A.~Sagnotti and M.~Taronna,
``String Lessons for Higher-Spin Interactions,''
Nucl. Phys. B \textbf{842} (2011), 299-361
%doi:10.1016/j.nuclphysb.2010.08.019
[arXiv:1006.5242 [hep-th]].

%\cite{Goddard:1973qh}
\bibitem{Goddard:1973qh}
P.~Goddard, J.~Goldstone, C.~Rebbi and C.~B.~Thorn,
``Quantum dynamics of a massless relativistic string,''
Nucl. Phys. B \textbf{56} (1973), 109-135
%doi:10.1016/0550-3213(73)90223-X

%\cite{DelGiudice:1971yjh}
\bibitem{DelGiudice:1971yjh}
E.~Del Giudice, P.~Di Vecchia and S.~Fubini,
``General properties of the dual resonance model,''
Annals Phys. \textbf{70} (1972), 378-398
%doi:10.1016/0003-4916(72)90272-2


%\cite{Curtright:1986di}
\bibitem{Curtright:1986di}
T.~L.~Curtright and C.~B.~Thorn,
``Symmetry Patterns in the Mass Spectra of Dual String Models,''
Nucl. Phys. B \textbf{274} (1986), 520-558
%doi:10.1016/0550-3213(86)90525-0

%\cite{Curtright:1986rr}
\bibitem{Curtright:1986rr}
T.~L.~Curtright, C.~B.~Thorn and J.~Goldstone,
``Spin Content of the Bosonic String,''
Phys. Lett. B \textbf{175} (1986), 47-52
%doi:10.1016/0370-2693(86)90329-1

%\cite{Hanany:2010da}
\bibitem{Hanany:2010da}
A.~Hanany, D.~Forcella and J.~Troost,
``The Covariant perturbative string spectrum,''
Nucl. Phys. B \textbf{846} (2011), 212-225
%doi:10.1016/j.nuclphysb.2011.01.002
[arXiv:1007.2622 [hep-th]].

%\cite{Bekaert:2006py}
\bibitem{Bekaert:2006py}
X.~Bekaert and N.~Boulanger,
``The unitary representations of the Poincar\'e group in any spacetime dimension,''
SciPost Phys. Lect. Notes \textbf{30} (2021), 1
%doi:10.21468/SciPostPhysLectNotes.30
[arXiv:hep-th/0611263 [hep-th]].

%\cite{Didenko:2014dwa}
\bibitem{Didenko:2014dwa}
V.~E.~Didenko and E.~D.~Skvortsov,
``Elements of Vasiliev Theory,''
Lect. Notes Phys. \textbf{1028} (2024), 269-456
%doi:10.1007/978-3-031-59656-8\_3
[arXiv:1401.2975 [hep-th]].

%\cite{Weinberg:1985tv}
\bibitem{Weinberg:1985tv}
S.~Weinberg,
``Coupling Constants and Vertex Functions in String Theories,''
Phys. Lett. B \textbf{156} (1985), 309-314
%doi:10.1016/0370-2693(85)91615-6

%\cite{Scherk:1974jj}
\bibitem{Scherk:1974jj}
J.~Scherk,
``An Introduction to the Theory of Dual Models and Strings,''
Rev. Mod. Phys. \textbf{47} (1975), 123-164
doi:10.1103/RevModPhys.47.123

%\cite{Henneaux:1986kp}
\bibitem{Henneaux:1986kp}
M.~Henneaux,
``Remarks on the Cohomology of the {BRS} Operator in String Theory,''
Phys. Lett. B \textbf{177} (1986), 35-38
%doi:10.1016/0370-2693(86)90009-2

\bibitem{Howe1}
R.~Howe, ``Remarks on classical invariant theory,'' Transactions of the
  American Mathematical Society {\bfseries 313} no.~2, (1989) 539--570.
 
\bibitem{Howe2}
R.~Howe, ``Transcending Classical Invariant Theory,'' J. Am. Math. Soc. \textbf{2} (1989) 535.

%\cite{Rowe:2012ym}
\bibitem{Rowe:2012ym}
D.~J.~Rowe, M.~J.~Carvalho and J.~Repka,
``Dual pairing of symmetry groups and dynamical groups in physics,''
Rev. Mod. Phys. \textbf{84} (2012), 711-757
%doi:10.1103/RevModPhys.84.711
[arXiv:1207.0148 [nucl-th]].

%\cite{Basile:2020gqi}
\bibitem{Basile:2020gqi}
T.~Basile, E.~Joung, K.~Mkrtchyan and M.~Mojaza,
``Dual Pair Correspondence in Physics: Oscillator Realizations and Representations,''
JHEP \textbf{09} (2020), 020
%doi:10.1007/JHEP09(2020)020
[arXiv:2006.07102 [hep-th]].  

%\cite{Brink:1976sc}
\bibitem{Brink:1976sc}
L.~Brink, P.~Di Vecchia and P.~S.~Howe,
``A Locally Supersymmetric and Reparametrization Invariant Action for the Spinning String,''
Phys. Lett. B \textbf{65} (1976), 471-474
%doi:10.1016/0370-2693(76)90445-7

%\cite{Polchinski:1998rr}
\bibitem{Polchinski:1998rr}
J.~Polchinski,
``String theory. Vol. 2: Superstring theory and beyond,''
Cambridge University Press, 2007,
%ISBN 978-0-511-25228-0, 978-0-521-63304-8, 978-0-521-67228-3
%doi:10.1017/CBO9780511618123

%\cite{Koh:1987hm}
\bibitem{Koh:1987hm}
I.~G.~Koh, W.~Troost and A.~Van Proeyen,
``Covariant Higher Spin Vertex Operators in the Ramond Sector,''
Nucl. Phys. B \textbf{292} (1987), 201-221
%doi:10.1016/0550-3213(87)90642-0

%\cite{Feng:2010yx}
\bibitem{Feng:2010yx}
W.~Z.~Feng, D.~Lust, O.~Schlotterer, S.~Stieberger and T.~R.~Taylor,
``Direct Production of Lightest Regge Resonances,''
Nucl. Phys. B \textbf{843} (2011), 570-601
%doi:10.1016/j.nuclphysb.2010.10.013
[arXiv:1007.5254 [hep-th]].

%\cite{Schlotterer:2010kk}
\bibitem{Schlotterer:2010kk}
O.~Schlotterer,
``Higher Spin Scattering in Superstring Theory,''
Nucl. Phys. B \textbf{849} (2011), 433-460
%doi:10.1016/j.nuclphysb.2011.03.026
[arXiv:1011.1235 [hep-th]].

%\cite{Lust:2023sfk}
\bibitem{Lust:2023sfk}
D.~L\"ust, C.~Markou, P.~Mazloumi and S.~Stieberger,
``A stringy massive double copy,''
JHEP \textbf{08} (2023), 193
%doi:10.1007/JHEP08(2023)193
[arXiv:2301.07110 [hep-th]].

%\cite{deRham:2010kj}
\bibitem{deRham:2010kj}
C.~de Rham, G.~Gabadadze and A.~J.~Tolley,
``Resummation of Massive Gravity,''
Phys. Rev. Lett. \textbf{106} (2011), 231101
%doi:10.1103/PhysRevLett.106.231101
[arXiv:1011.1232 [hep-th]].

%\cite{Hassan:2011zd}
\bibitem{Hassan:2011zd}
S.~F.~Hassan and R.~A.~Rosen,
``Bimetric Gravity from Ghost-free Massive Gravity,''
JHEP \textbf{02} (2012), 126
%doi:10.1007/JHEP02(2012)126
[arXiv:1109.3515 [hep-th]].

%\cite{Hassan:2011ea}
\bibitem{Hassan:2011ea}
S.~F.~Hassan and R.~A.~Rosen,
``Confirmation of the Secondary Constraint and Absence of Ghost in Massive Gravity and Bimetric Gravity,''
JHEP \textbf{04} (2012), 123
%doi:10.1007/JHEP04(2012)123
[arXiv:1111.2070 [hep-th]].

%\cite{Lust:2021jps}
\bibitem{Lust:2021jps}
D.~L\"ust, C.~Markou, P.~Mazloumi and S.~Stieberger,
``Extracting bigravity from string theory,''
JHEP \textbf{12} (2021), 220
%doi:10.1007/JHEP12(2021)220
[arXiv:2106.04614 [hep-th]].

%\cite{Guevara:2018wpp}
\bibitem{Guevara:2018wpp}
A.~Guevara, A.~Ochirov and J.~Vines,
``Scattering of Spinning Black Holes from Exponentiated Soft Factors,''
JHEP \textbf{09} (2019), 056
%doi:10.1007/JHEP09(2019)056
[arXiv:1812.06895 [hep-th]].

%\cite{Maybee:2019jus}
\bibitem{Maybee:2019jus}
B.~Maybee, D.~O'Connell and J.~Vines,
``Observables and amplitudes for spinning particles and black holes,''
JHEP \textbf{12} (2019), 156
%doi:10.1007/JHEP12(2019)156
[arXiv:1906.09260 [hep-th]].

%\cite{Cangemi:2022bew}
\bibitem{Cangemi:2022bew}
L.~Cangemi, M.~Chiodaroli, H.~Johansson, A.~Ochirov, P.~Pichini and E.~Skvortsov,
``Kerr Black Holes From Massive Higher-Spin Gauge Symmetry,''
Phys. Rev. Lett. \textbf{131} (2023) no.22, 221401
%doi:10.1103/PhysRevLett.131.221401
[arXiv:2212.06120 [hep-th]].

%\cite{Cangemi:2022abk}
\bibitem{Cangemi:2022abk}
L.~Cangemi and P.~Pichini,
``Classical limit of higher-spin string amplitudes,''
JHEP \textbf{06} (2023), 167
%doi:10.1007/JHEP06(2023)167
[arXiv:2207.03947 [hep-th]].

%\cite{Callan:1996dv}
\bibitem{Callan:1996dv}
C.~G.~Callan and J.~M.~Maldacena,
``D-brane approach to black hole quantum mechanics,''
Nucl. Phys. B \textbf{472} (1996), 591-610
%doi:10.1016/0550-3213(96)00225-8
[arXiv:hep-th/9602043 [hep-th]].

%\cite{Dhar:1996vu}
\bibitem{Dhar:1996vu}
A.~Dhar, G.~Mandal and S.~R.~Wadia,
``Absorption versus decay of black holes in string theory and T symmetry,''
Phys. Lett. B \textbf{388} (1996), 51-59
%doi:10.1016/0370-2693(96)01127-6
[arXiv:hep-th/9605234 [hep-th]].

%\cite{Das:1996wn}
\bibitem{Das:1996wn}
S.~R.~Das and S.~D.~Mathur,
``Comparing decay rates for black holes and D-branes,''
Nucl. Phys. B \textbf{478} (1996), 561-576
%doi:10.1016/0550-3213(96)00453-1
[arXiv:hep-th/9606185 [hep-th]].

%\cite{Maldacena:1996ix}
\bibitem{Maldacena:1996ix}
J.~M.~Maldacena and A.~Strominger,
``Black hole grey body factors and d-brane spectroscopy,''
Phys. Rev. D \textbf{55} (1997), 861-870
%doi:10.1103/PhysRevD.55.861
[arXiv:hep-th/9609026 [hep-th]].

%\cite{Hashimoto:1996bf}
\bibitem{Hashimoto:1996bf}
A.~Hashimoto and I.~R.~Klebanov,
``Scattering of strings from D-branes,''
Nucl. Phys. B Proc. Suppl. \textbf{55} (1997), 118-133
%doi:10.1016/S0920-5632(97)00074-1
[arXiv:hep-th/9611214 [hep-th]].

%\cite{Bianchi:2017sds}
\bibitem{Bianchi:2017sds}
M.~Bianchi, D.~Consoli and J.~F.~Morales,
``Probing Fuzzballs with Particles, Waves and Strings,''
JHEP \textbf{06} (2018), 157
%doi:10.1007/JHEP06(2018)157
[arXiv:1711.10287 [hep-th]].

%\cite{Bianchi:2018kzy}
\bibitem{Bianchi:2018kzy}
M.~Bianchi, D.~Consoli, A.~Grillo and J.~F.~Morales,
``The dark side of fuzzball geometries,''
JHEP \textbf{05} (2019), 126
%doi:10.1007/JHEP05(2019)126
[arXiv:1811.02397 [hep-th]].

%\cite{Gross:2021gsj}
\bibitem{Gross:2021gsj}
D.~J.~Gross and V.~Rosenhaus,
``Chaotic scattering of highly excited strings,''
JHEP \textbf{05} (2021), 048
%doi:10.1007/JHEP05(2021)048
[arXiv:2103.15301 [hep-th]].

%\cite{Bianchi:2022mhs}
\bibitem{Bianchi:2022mhs}
M.~Bianchi, M.~Firrotta, J.~Sonnenschein and D.~Weissman,
``Measure for Chaotic Scattering Amplitudes,''
Phys. Rev. Lett. \textbf{129} (2022) no.26, 261601
%doi:10.1103/PhysRevLett.129.261601
[arXiv:2207.13112 [hep-th]].


%\cite{Pesando:2024lqa}
\bibitem{Pesando:2024lqa}
I.~Pesando,
``The bosonic string spectrum and the explicit states up to level 10 from the lightcone and the chaotic behavior of certain string amplitudes,''
Eur. Phys. J. C \textbf{85} (2025) no.4, 371
%doi:10.1140/epjc/s10052-025-13955-y
[arXiv:2405.09987 [hep-th]].

%\cite{Bianchi:2003wx}
\bibitem{Bianchi:2003wx}
M.~Bianchi, J.~F.~Morales and H.~Samtleben,
``On stringy AdS(5) x S**5 and higher spin holography,''
JHEP \textbf{07} (2003), 062
%doi:10.1088/1126-6708/2003/07/062
[arXiv:hep-th/0305052 [hep-th]].

%\cite{Beisert:2003te}
\bibitem{Beisert:2003te}
N.~Beisert, M.~Bianchi, J.~F.~Morales and H.~Samtleben,
``On the spectrum of AdS / CFT beyond supergravity,''
JHEP \textbf{02} (2004), 001
%doi:10.1088/1126-6708/2004/02/001
[arXiv:hep-th/0310292 [hep-th]].

%\cite{Beisert:2004di}
\bibitem{Beisert:2004di}
N.~Beisert, M.~Bianchi, J.~F.~Morales and H.~Samtleben,
``Higher spin symmetry and N=4 SYM,''
JHEP \textbf{07} (2004), 058
%doi:10.1088/1126-6708/2004/07/058
[arXiv:hep-th/0405057 [hep-th]].

%\cite{Gaberdiel:2018rqv}
\bibitem{Gaberdiel:2018rqv}
M.~R.~Gaberdiel and R.~Gopakumar,
``Tensionless string spectra on AdS$_{3}$,''
JHEP \textbf{05} (2018), 085
%doi:10.1007/JHEP05(2018)085
[arXiv:1803.04423 [hep-th]].

%\cite{Sundborg:2000wp}
\bibitem{Sundborg:2000wp}
B.~Sundborg,
``Stringy gravity, interacting tensionless strings and massless higher spins,''
Nucl. Phys. B Proc. Suppl. \textbf{102} (2001), 113-119
%doi:10.1016/S0920-5632(01)01545-6
[arXiv:hep-th/0103247 [hep-th]].

%\cite{Taronna:2011kt}
\bibitem{Taronna:2011kt}
M.~Taronna,
``Higher-Spin Interactions: four-point functions and beyond,''
JHEP \textbf{04} (2012), 029
%doi:10.1007/JHEP04(2012)029
[arXiv:1107.5843 [hep-th]].

\end{thebibliography}
\end{document}